\def\hetetwo{{\it HETE-2}\ }
\def\hetetwonosp{{\it HETE-2}}

\def\bsax{{\it Beppo}SAX\ }
\def\bsaxnosp{{\it Beppo}SAX}
\def\eop{E^{\rm obs}_{\rm peak}}
\def\eiso{E_{\rm iso}}

\def\ep{E_{\rm peak}}
\def\davg{\langle D \rangle}
\def\dnavg{\langle D(0) \rangle}
\def\egi{E^{\rm inf}_{\gamma}}
\def\egt{E^{\rm true}_{\gamma}}

\def\ojet{\Omega_{0}}
\def\thetav{\theta_{\rm v}}
\def\thetan{\theta_{0}}
\def\thetabr{\theta_{\rm br}}
\def\egamma{E_{\gamma}}
\def\epei{\ep \propto \eiso^{1/2}}
\def\epeioff{\ep \propto \eiso^{1/3}}
\def\epeg{\ep \propto \egamma^{\beta}}

\documentclass{emulateapj}

\newcommand{\beq}{\begin{equation}}
\newcommand{\eeq}{\end{equation}}
\newcommand{\bea}{\begin{eqnarray}}
\newcommand{\eea}{\end{eqnarray}}

\shorttitle{Off-Jet Emission}
\shortauthors{Donaghy}

\begin{document}

\title{The Importance of Off-Jet Relativistic Kinematics in Gamma-Ray 
Burst Jet Models}
\author{T. Q. Donaghy}
\affil{Department of Physics, University of Chicago, 
5640 S. Ellis Avenue, Chicago, IL 60637}
\email{quinn@oddjob.uchicago.edu}

\begin{abstract}

Gamma-Ray Bursts (GRBs) are widely thought to originate from collimated
jets of material moving at relativistic velocities.  Emission from such
a jet should be visible even when viewed from outside the angle of
collimation.  I summarize recent work on the special relativistic 
transformation of the burst quantities $\eiso$ and $\ep$ as a function
of viewing angle, where $\eiso$ is the isotropic-equivalent energy of
the burst and $\ep$ is the peak of the burst spectrum in the power $\nu
F_{\nu}$.  The formulae resulting from this work serve as input for a
Monte Carlo population synthesis method, with which I investigate the
importance of off-jet relativistic kinematics as an explanation for a
class of GRBs termed ``X-ray Flashes'' (XRFs).  I do this in the context
of several top-hat shaped variable opening-angle jet models.  I find
that such models predict a large population of off-jet bursts that are
observable and that lie away from the $\epei$ relation.  The predicted
burst populations are not seen in current datasets.  I investigate the
effect of the bulk $\gamma$ value upon the properties of this population
of off-jet bursts, as well as the effect of including an
$\ojet$-$\egamma$ correlation to jointly fit the $\epei$ and $\epeg$
relations, where $\ojet$ is the opening solid angle of the GRB jet.  I
find that the XRFs seen by \hetetwo and \bsax cannot be easily explained
as classical GRBs viewed off-jet.  I also find that an inverse
correlation between $\gamma$ and $\ojet$ has the effect of greatly
reducing the visibility of off-jet events.  Therefore, unless $\gamma >
300$ for all bursts or unless there is a strong inverse correlation
between $\gamma$ and $\ojet$, top-hat variable opening-angle jet models
produce a significant population of bursts away from the $\epei$ and
$\epeg$ relations, in contradiction of current observations.

\end{abstract}


\section{Introduction}
\label{cha:intro}

The importance of collimated jets in GRBs was highlighted by the
extremely large isotropic-equivalent energies ($\eiso$) of very luminous
events like GRB 971214 \citep{kulkarni1998} and GRB 990123
\citep{kulkarni1999} and by the observation of breaks in afterglow
light-curves \citep{rhoads1997,sph1999,harrison1999}.  \cite{frail2001}
and \cite{bloom2003} corrected the isotropic-equivalent energies by the
beaming fraction obtained from the jet break time in afterglow
light-curves and found that the values of the energy released in
$\gamma$-rays ($\egamma$) were tightly clustered around $10^{51}$ ergs. 
Recently \cite{ghirlanda2004} have shown that a tight correlation exists
between $\egamma$ and the peak of the $\nu F_{\nu}$ spectrum in the
rest-frame, $\ep$.  Recent results by \cite{sakamoto2005a} obtained from
\hetetwo \citep{ricker2003} observations have shown that XRFs
\citep{heise2000,kippen2002}, X-ray-rich GRBs and GRBs lie along a
continuum of properties and that XRFs with known redshift extend the
$\epei$ relation predicted by \cite{lloyd-ronning2000} and found by
\cite{amati2002} to over five orders of magnitude in $\eiso$
\citep{lamb2005}.

Relativistic kinematics implies that even a ``top-hat''-shaped jet will
be visible when viewed outside its angle of collimation; i.e., off-jet 
\citep{ioka2001}.  \cite{yamazaki2002,yamazaki2003} used this fact to
construct a model where XRFs are simply classical GRBs viewed at an
angle $\thetav > \thetan$, where $\thetan$ is the half-opening angle of
the jet and $\thetav$ is the angle between the axis of the jet and the
line-of-sight.  The authors showed that such a model could reproduce
many of the observed characteristics of XRFs.  \cite{yamazaki2004a}
showed that in such a model, the distribution of both on- and off-jet
observed bursts was roughly consistent with the $\epei$ relation.

In this paper, I use the population synthesis method developed by
\cite{ldg2005} and incorporate the relativistic emission profiles
calculated by \cite{graziani2005}, to predict the global properties of
bursts localized by \hetetwo and \bsaxnosp.  I consider the possibility
that the XRFs observed by \hetetwo and \bsax are primarily regular GRBs
observed off-jet \citep{yamazaki2004a} and show that it is difficult to
account for the observed properties of XRFs in this model.  However,
since the effect of special relativity on off-jet emission must exist, I
seek to understand its relative importance in the context of current
models of GRB jets.  I revisit the top-hat variable opening-angle
(THVOA) jet model put forward in \cite{ldg2005}, now including the
effects of relativistic kinematics on off-jet emission.  I present
results for several models which explore various regions of the
parameter space in $\gamma$, $\egamma$ and $\thetan$.

For this paper, I only consider the effect of relativistic kinematics on
off-jet emission\footnote{In this paper I use the terms``off-jet
emission'' or ``off-jet relativistic kinematics'', rather than
``off-axis beaming'', to emphasize that such emission is a direct
consequence of special relativity and that it is primarily important
beyond the edge of the jet.} from uniform or ``top-hat'' jets; we will
consider the effects on Fisher-shaped \citep{dlg2005a} and
Gaussian-shaped \citep{zhang2004} jets in a future publication
\citep{dlg2005b}.  I describe my population synthesis method in \S 2 and
present the results for various models in \S 3.  I discuss the results
in \S 4 and draw some conclusions in \S 5.  Preliminary results were
reported in \cite{donaghy2005a}.

\section{Method}

\subsection{Off-Jet Relativistic Kinematic Formulae}

Relativistic kinematics causes frequencies in the rest frame of the jet
to appear Doppler shifted by a factor  $\delta = \gamma (1-\beta
\cos\theta)$, where $\beta$ is the bulk velocity of the jet and $\theta$
is the angle between the direction of motion and the source frame
observer.  The simulations I describe chiefly deal with the kinematic
transformation of two important burst quantities, $\eiso$ and $\ep$, as
a function of viewing angle, $\thetav$.  In the simplest ``toy'' model,
these quantities transform as $\ep \propto \delta^{-1}$ and $\eiso
\propto \delta^{-3}$, from which arises the relation $\epeioff$
\citep{yamazaki2002}.  In the more complete model of
\cite{graziani2005}, this relation is satisfied only in the limit
$\thetav \gg \thetan$.

The complete relativistic kinematic expressions involve convolution of
the Doppler function and the intrinsic profile of the jet.  For an
arbitrary smooth profile, an efficient algorithm exists to calculate the
profiles; for the case I am interested in, the uniform or ``top-hat''
profile, a closed analytic expression can be given.  The formulae are
derived in \cite{graziani2005}, and I summarize them below.  The current
model differs slightly from that used by \cite{yamazaki2004a} in that we
consider steady-state emission, rather than the evolution of burst
properties due to time-of-flight effects.  The model therefore applies
to burst-averaged data products like $\eiso$ and $\ep$.

The observed isotropic-equivalent energy, $\eiso$, of the jet as a
function of $\thetav$ is given by, 
\beq
\eiso = \frac{\egt}{2\beta\gamma^{4} (1-\cos\thetan)}
	\left[ f(\beta-\cos\thetav) - f(\beta\cos\thetan-\cos\thetav) 
	\right], 
\label{eqn_eiso}\eeq
where
\beq
f(z) = \frac{\gamma^{2}(2\gamma^{2}-1)z^{3} + (3\gamma^{2}\sin^{2}\thetav)z
	+ 2\cos\thetav\sin^{2}\thetav}{(z^{2} +
	\gamma^{-2}\sin^{2}\thetav)^{3/2}},
\eeq
and $\egt$ is the total energy emitted by the jet in gamma rays and
serves as the energy scale for the emission profile.

\begin{figure*}[htb]
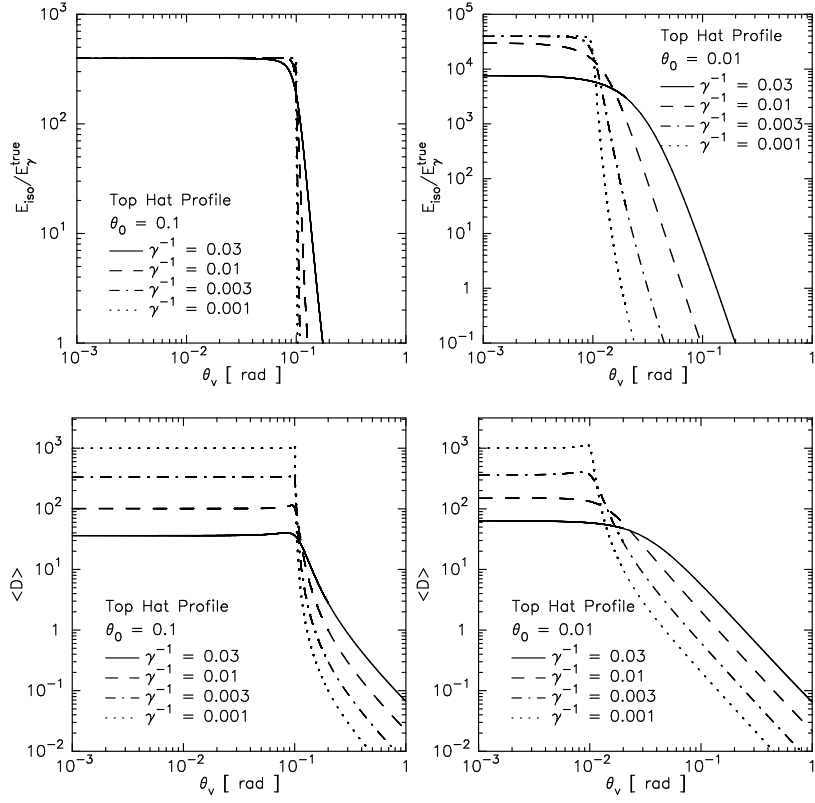

\begin{center}
\resizebox{5.3cm}{!}{\includegraphics{f1a.eps}}
\resizebox{5.3cm}{!}{\includegraphics{f1b.eps}}
\end{center}
\begin{center}
\resizebox{5.3cm}{!}{\includegraphics{f1c.eps}}
\resizebox{5.3cm}{!}{\includegraphics{f1d.eps}}
\end{center}
\caption{Emission profiles (top row) and average Doppler shift (bottom
row) as a function of viewing angle, for a range of $\gamma$ and
$\thetan =0.1$ rad (left) and $\thetan =0.01$ rad (right).  From
\cite{graziani2005}.}
\label{carlo_form}
\end{figure*}

\begin{figure*}[t]
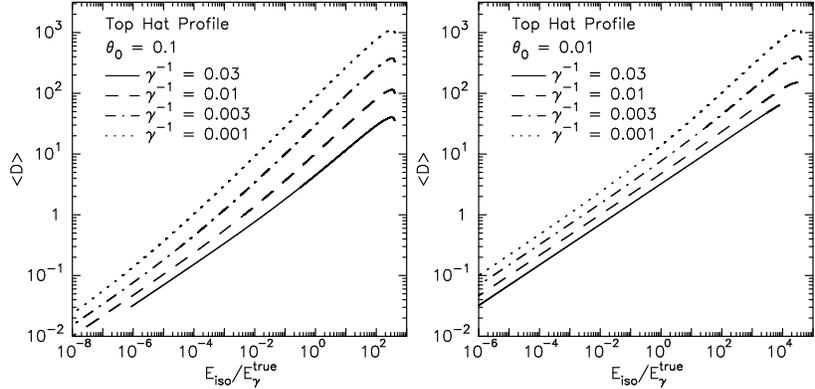

\begin{center}
\resizebox{5.3cm}{!}{\includegraphics{f2a.eps}}
\resizebox{5.3cm}{!}{\includegraphics{f2b.eps}}
\end{center}
\caption{Trajectories in the [$\eiso$,$\davg$]-plane (``Amati plot''),
moving from right to left as $\thetav$ increases, for $\thetan =0.1$ rad
(left) and $\thetan =0.01$ rad (right).  The axes are scaled so that a
$45^{\circ}$ line corresponds to a slope of $1/2$.  From
\cite{graziani2005}.}
\label{carlo_traj}
\end{figure*}

The transformation of $\ep$ as a function of $\thetav$ is slightly more
complicated.  The detailed physics underlying the prompt emission of
gamma-ray bursts is not yet well understood.  In particular, an
explanation of the prompt emission spectrum (apparently universally
parametrized by the Band function \citep{band1993}) is currently
lacking.  The observed spectrum (including the value of $\ep$) is almost
certainly due to superpositions of emission from different regions on
the jet, convolved with relativistic kinematic effects.  A detailed
explanation of how this forms a Band spectrum is beyond the scope of
this paper.  Instead we calculate the average Doppler shift across the
jet as a proxy for $\ep$.  The average shift, $\davg$, is given by,
\beq
\davg = \gamma^{-1} 
	\frac{f(\beta-\cos\thetav) - f(\beta\cos\thetan-\cos\thetav)}
	{g(\beta-\cos\thetav) - g(\beta\cos\thetan-\cos\thetav)},
\eeq
where
\beq
g(z) = \frac{2\gamma^{2}z + 2\cos\thetav}
	{(z^{2} + \gamma^{-2}\sin^{2}\thetav)^{1/2}}.
\eeq
$\ep$ is related to $\davg$ via,
\beq
\ep = \davg \times \ep^{\rm (rest)},
\eeq
where $\ep^{\rm (rest)}$ is the (unknown) peak energy of the burst
spectrum in the rest frame.  We remove this unknown normalization by
requiring all bursts to obey the $\epei$ relation at the center of the
jet, thereby fixing this normalization to that of $\eiso$.  Thus,
\beq
\ep = \frac{\langle D(\thetav) \rangle}{\dnavg} \cdot
	\ep^{(\rm on)} 
	= \frac{\langle D(\thetav) \rangle}{\dnavg} \cdot
	C_{A} \cdot \left( \eiso^{(\rm on)}/ E_{A} \right)^{0.5},
\label{eqn_ep}\eeq
where $\eiso^{(\rm on)}$ is simply Equation \ref{eqn_eiso} evaluated at
$\thetav=0$.  Figure \ref{carlo_form} shows $\eiso$ and $\davg$ plotted
as functions of $\thetav$ for various values of $\thetan$ and $\gamma$,
and Figure \ref{carlo_traj} shows the corresponding trajectories in the
[$\eiso$,$\davg$]-plane as $\thetav$ increases away from the jet axis.

\subsection{Monte Carlo Simulations}

The population synthesis Monte Carlo simulations described in this paper
follow the method presented in \cite{ldg2005}.  Beginning in the
rest-frame of each burst I specify $\thetan$, $\egt$ and $\gamma$ by
drawing from various input distributions (depending on the model used). 
A value for $\thetav$ is drawn from the distribution $dN = \sin(\thetav)
\; d\thetav$.  I then calculate $\eiso$ and $\ep$ from Equations
\ref{eqn_eiso} and \ref{eqn_ep}.

I also introduce two Gaussian smearing functions to add a stochastic
element to the simulations.  I draw values for the coefficient $C_{A}$
in Equation \ref{eqn_ep} from a narrow lognormal distribution to model
the observed width of the $\epei$ relation.  Another narrow lognormal
distribution is used to generate a timescale for each burst that
converts fluences to peak fluxes.  For both of these Gaussians I use the
same parameters as in \cite{ldg2005}.

I then draw redshift values from a model for the star-formation rate
\citep{rr2001}, transform the burst quantities to the observer-frame and
construct a Band spectrum (assuming $\alpha=-1$ and $\beta=-2.5$). 
Using the observer-frame Band spectrum I can calculate photon and energy
fluences and peak fluxes in any desired passband.  By comparing with the
peak photon flux thresholds as described by \cite{band2003}, I determine
if a burst would be detected by a given instrument.  In this work, I
primarily employ the detector thresholds from the WXM on \hetetwonosp,
scaled to include triggers on timescales up to $5$ sec.

One change in the method from \cite{ldg2005} is necessarily the
treatment of off-jet events.  In that paper, simulation of all off-jet
events was bypassed by drawing from a power-law distribution in $\ojet$
with index $\delta_{\rm sim} = \delta_{\rm true} - 1$; as discussed in
that work, this is equivalent to working in the $\gamma \to \infty$
limit.  In that paper, we concluded that the data requested a model with
approximately equal numbers of bursts per decade in all observed burst
quantities, corresponding to $\delta_{\rm sim} = 1$.  Equivalently, in
this paper I draw from a power-law distribution in $\ojet$ with index
$\delta_{\rm true} = 2$, and simulate the distribution of viewing
angles, $\thetav$, to determine which off-jet events are detected.  

The vast majority of jets are observed from large viewing angles and
therefore are extremely faint and are not detected.  In order to
increase the percentage of detected bursts in a Monte Carlo sample of
$50,000$ events, I do not simulate bursts that lie in regions of
parameter space that produce faint bursts; i.e. bursts with values of 
$\thetav$ which are large compared to $\thetan$.  For each model, I plot
bursts in the [$\thetan$,$\thetav$]-plane, showing the outline of the
truncated region.  The models in \cite{ldg2005} correspond to removing
all bursts above the line $\thetan = \thetav$.

\subsection{Comparison to Burst Data}

To test the viability of various jet models, I compare my results
against several available datasets.  Considering the sample of bursts
with known redshifts (localized by \bsaxnosp, \hetetwo or other
detectors) there are two correlations of interest between source frame
quantities.

\cite{amati2002} reported a correlation between $\ep$ and $\eiso$
\beq
\ep = C_{A} \left( \frac{\eiso}{E_{A}} \right)^{\alpha},
\label{eqn_amati}\eeq
that has been confirmed and extended to over five orders of magnitude in
$\eiso$ \citep{sakamoto2004,sakamoto2005a,lamb2005}.  I work with the
best-fit value of $C_{A} = 89.1 \pm 8.2$ keV (fixing $\alpha = 0.5$ and
$E_{A} = 10^{52}$ ergs) from \cite{ldg2005}.  For each realization, I
draw a value for $C_{A}$ from a lognormal distribution with a width of
$0.13$ decades.

Recent works \citep{nakar2004a,band2005} have claimed that large
percentages of BATSE bursts are incompatible with the $\epei$ relation
(but see \cite{ghirlanda2005a}, \cite{bosnjak2005} and
\cite{lamb2005}).  Since bursts seen away from the $\epei$ relation may
be a signature of off-jet events, comparison with this relation is an
important test for these models.

Using bursts with known redshift and well-measured jet-break times,
\cite{ghirlanda2004} have found a second relation
\beq
\ep = C_{G} \left( \frac{\egamma}{E_{G}} \right)^{\beta},
\label{eqn_ggl}\eeq
with current best-fit values of $\beta = 0.69 \pm 0.04$, $C_{G} = 250
\pm 100$ keV and $E_{G} = 3.8 \times 10^{50}$ ergs
\citep{ghirlanda2005b}.  In \S 3.3 I describe the observational
implications of models that satisfy both the $\epei$ and $\epeg$
relations.

I define the value of $\egamma$ measured using the method of
\cite{frail2001} to be $\egi = \eiso (1-\cos\thetabr)$, where $\thetabr
= \max(\thetav,\thetan)$.  The presence of off-jet emission (or any
non-uniform profile) implies $\egi \ne \egt$; in fact, $\egi$ varies
with $\thetav$ for a given jet, while $\egt$ is an intrinsic property of
the jet.

The burst properties presented in the source frame ($\eiso$ or $\egi$
against $\ep$, or cumulative distributions of these quantities) are
essentially those presented by \cite{ghirlanda2004}, augmented where
appropriate with events compiled by \cite{friedman2004}, more recent
fits to \hetetwo data from \cite{sakamoto2005a}, and a few recent Swift
bursts with fits reported in GCN Circulars
\citep{golenetskii2005a,golenetskii2005b,golenetskii2005c}.

\begin{figure}[htb]
\begin{center}
\rotatebox{270}{\resizebox{6cm}{!}{\includegraphics{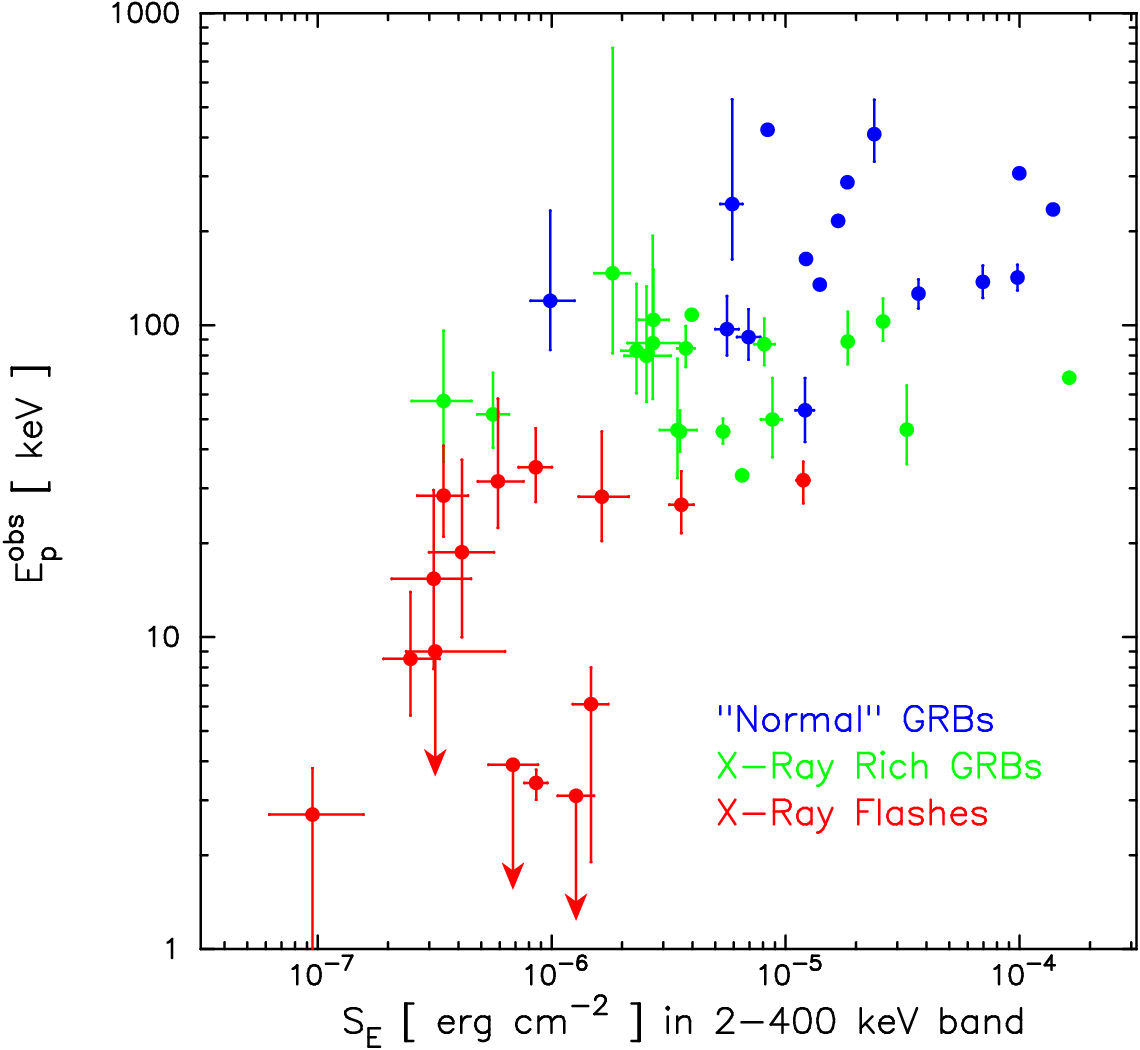}}}
\rotatebox{270}{\resizebox{6cm}{!}{\includegraphics{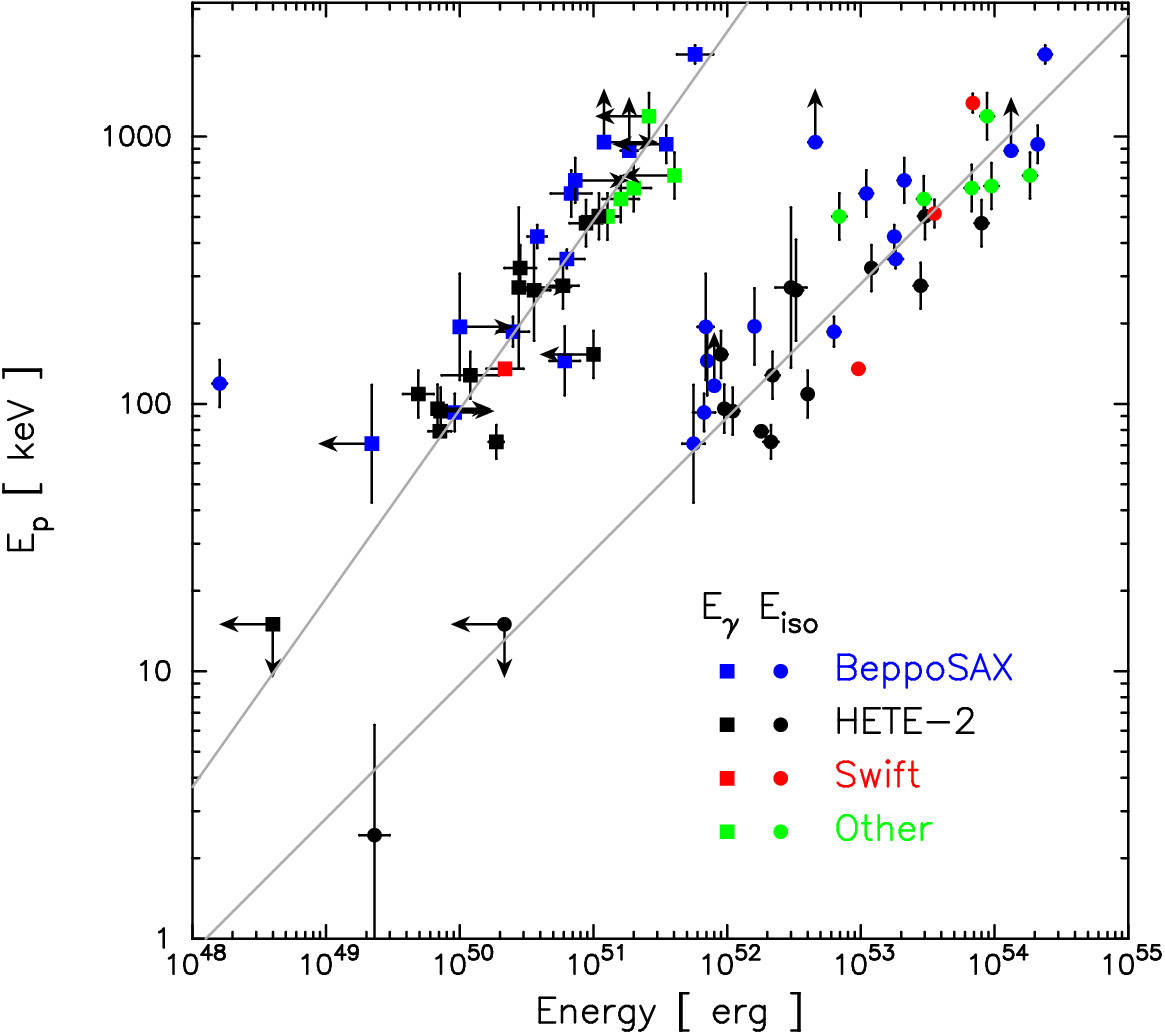}}}
\end{center}
\caption{Top panel shows the sample of bursts localized by the WXM on
\hetetwo in the [$S_{E}$(2-400 keV),$\eop$]-plane, broken down into
XRFs, X-ray-rich GRBs and GRBs (data taken from \cite{sakamoto2005a} and
\cite{barraud2003}).  Bottom panel shows the $\epei$ and $\epeg$
relations in the source frame (data taken from \cite{ghirlanda2004}, 
\cite{friedman2004}, \cite{sakamoto2005a} and recent Swift bursts).}
\label{datasets}
\end{figure}

I also consider the larger sample of \hetetwo localized bursts with and
without known redshifts \citep{barraud2003,sakamoto2005a}.  This sample
also shows a prominent hardness-intensity correlation, although it is
broader than the source-frame correlation.  This sample has the
advantage of having many more XRFs than the sample with known
redshifts.  Figure \ref{datasets} shows both the observer frame and
source frame datasets, with the relevant correlations.

\section{Results}

Here I explore the relative importance of off-jet relativistic
kinematics for six top-hat variable opening-angle jet models.  In what
follows, bursts that are detected by the WXM for which $\thetav <
\thetan$ are shown as blue dots (on-jet events), while detected bursts
for which $\thetav > \thetan$ are shown as green dots (off-jet events). 
Bursts in the simulation which are not detected are shown as red dots.

\subsection{\cite{yamazaki2004a} Model}

In the first model I consider (Y04), I adopt the parameters from
\cite{yamazaki2004a}.  This model attempts to explain classical GRBs in
terms of the variation of jet opening-angles, while XRFs are interpreted
as classical GRBs viewed off-jet.  The Y04 model assumes $\gamma=100$
and draws $\thetan$ values from a power-law distribution given by
$f_{0}\,d\thetan \propto \thetan^{-2}\,d\thetan$, defined from $0.3$ to
$0.03$ rad.  $\egt$ is drawn from a narrow lognormal distribution
centered on $1.2 \times 10^{51}$ erg.  

\begin{figure*}[htb]
\begin{center}
\rotatebox{270}{\resizebox{5.3cm}{!}{\includegraphics{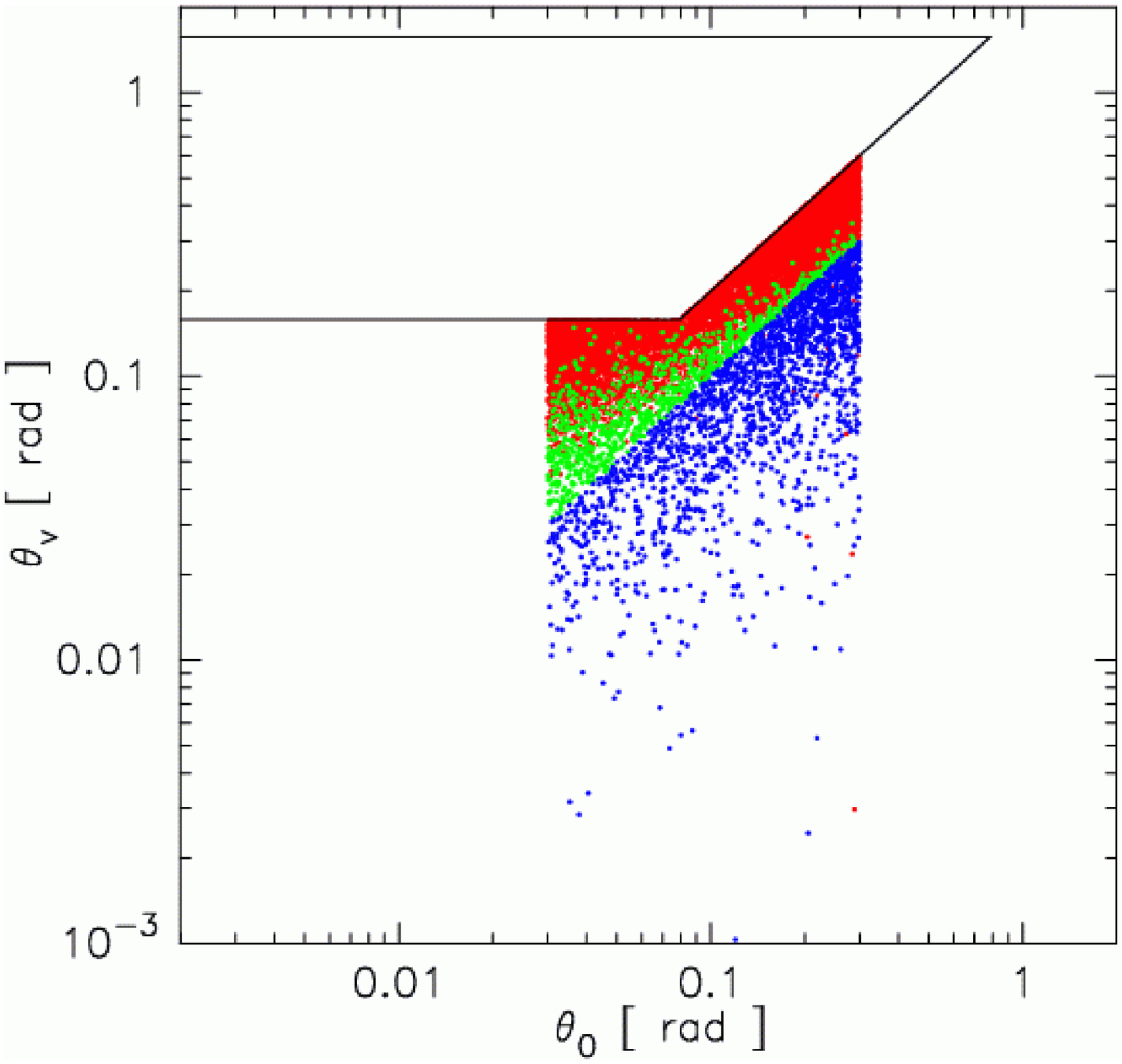}}}
\rotatebox{270}{\resizebox{5.3cm}{!}{\includegraphics{f4b.eps}}}
\rotatebox{270}{\resizebox{5.3cm}{!}{\includegraphics{f4c.eps}}}
\end{center}
\caption{Model Y04, which uses the parameters of \cite{yamazaki2004a},
where GRBs are seen on-jet and XRFs are explained by off-jet
relativistic kinematics.  Bursts detected on-jet (blue), off-jet (green)
and non-detected bursts (red) are shown in the
[$\thetan$,$\thetav$]-plane (left), [$\eiso$,$\ep$]-plane (center) and
the [$\eop$,$S_{E}$(2-400 keV)]-plane (right), with observed data from
\bsax and \hetetwo overplotted.}
\label{fig_y04}
\end{figure*}

The upper-right panel of Figure \ref{fig_y04} shows that the standard
$\egamma$ value and a narrow range of opening angles is sufficient to
explain the population of classical GRBs.  By construction, the on-jet
events follow the $\epei$ relation.   Off-jet emission from similar
bursts viewed at much larger $\thetav$ accounts for the population of
green off-jet points that lie above the $\epei$ relation.  Off-jet events
account for $34$\% of the total detected bursts (see Table
\ref{table_stats}).  For different values of $\thetav$, these bursts
generally move along trajectories in the [$\eiso$,$\ep$]-plane that have
a flatter slope than the $\epei$ relation (see Figure \ref{carlo_traj}).
The observed off-jet bursts (green points) are consistent with only a few
observed bursts, and for the most part, represent a population of events
not seen by current instruments.  

In particular, the middle and right panels of Figure \ref{fig_y04} show
that the \hetetwo XRFs are not easily explained as classical GRBs viewed
off-jet, as this model posits.  The two XRFs with known redshifts lie
along the $\epei$ relation, and furthermore, the larger sample of
\hetetwo XRFs without known redshifts do not fall in the region of the
[$\eop$,$S_{\rm E}$]-plane expected for this model (they lie at lower,
rather than higher, $\eop$ values for a given $S_{\rm E}$).

\subsection{Top-Hat Variable Opening-Angle Jet Models}

The next two models seek to explain both GRBs and XRFs by a wide
distribution of jet opening-angles (see \cite{ldg2005} for more details
and discussion).  These models generate XRFs that obey the $\epei$
relation by extending the range of possible jet opening solid-angles to
cover five orders of magnitude.  Hence, XRFs that obey the $\epei$
relation are bursts that are seen on-jet, but have larger jet
opening-angles.  Here I add the presence of off-jet relativistic
kinematics to this picture.   These models generate a significant
population of off-jet events, although increasing $\gamma$ reduces the
fraction of off-jet bursts in the observed sample.  

\begin{figure*}[htb]
\begin{center}
\rotatebox{270}{\resizebox{5.3cm}{!}{\includegraphics{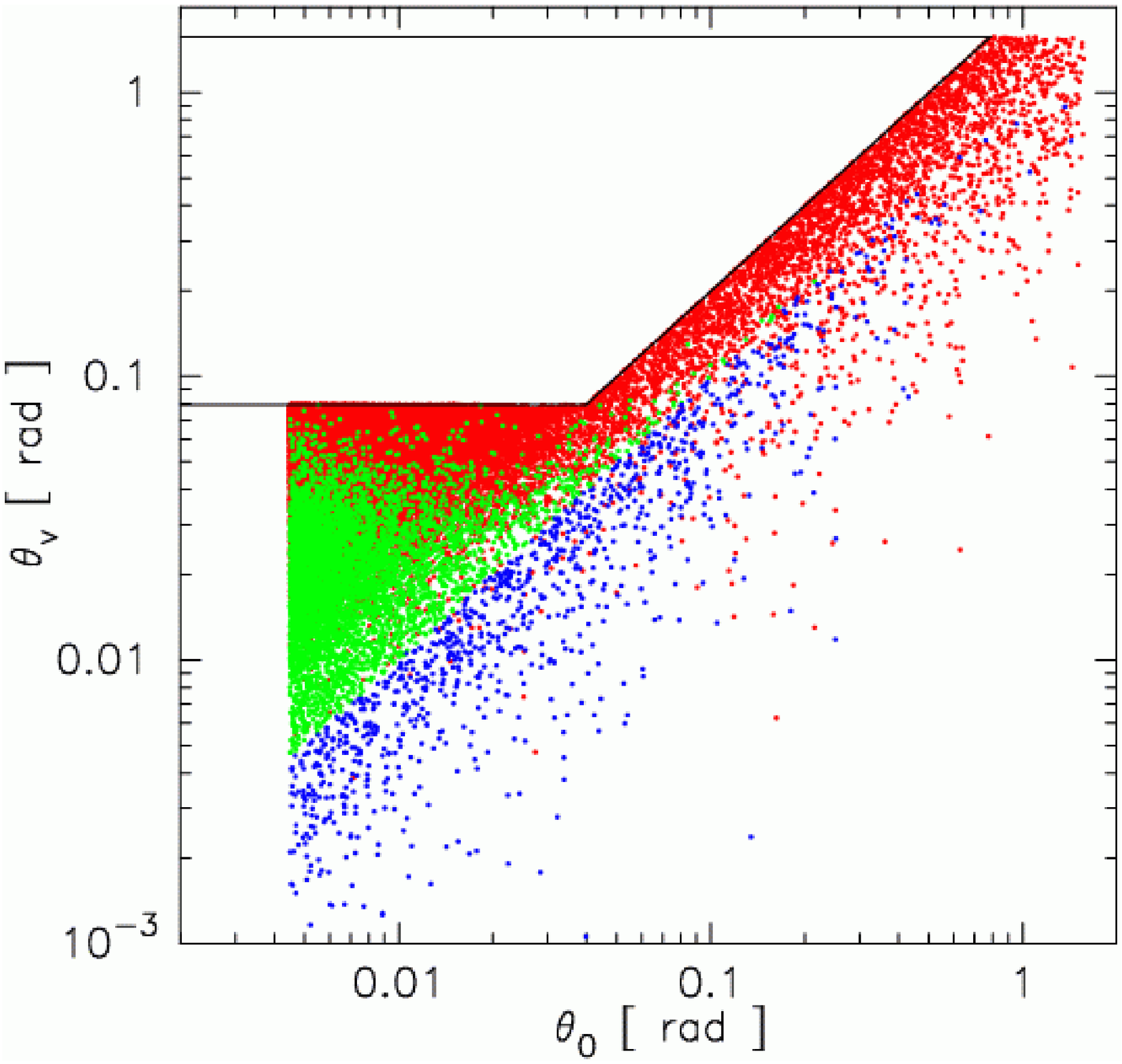}}}
\rotatebox{270}{\resizebox{5.3cm}{!}{\includegraphics{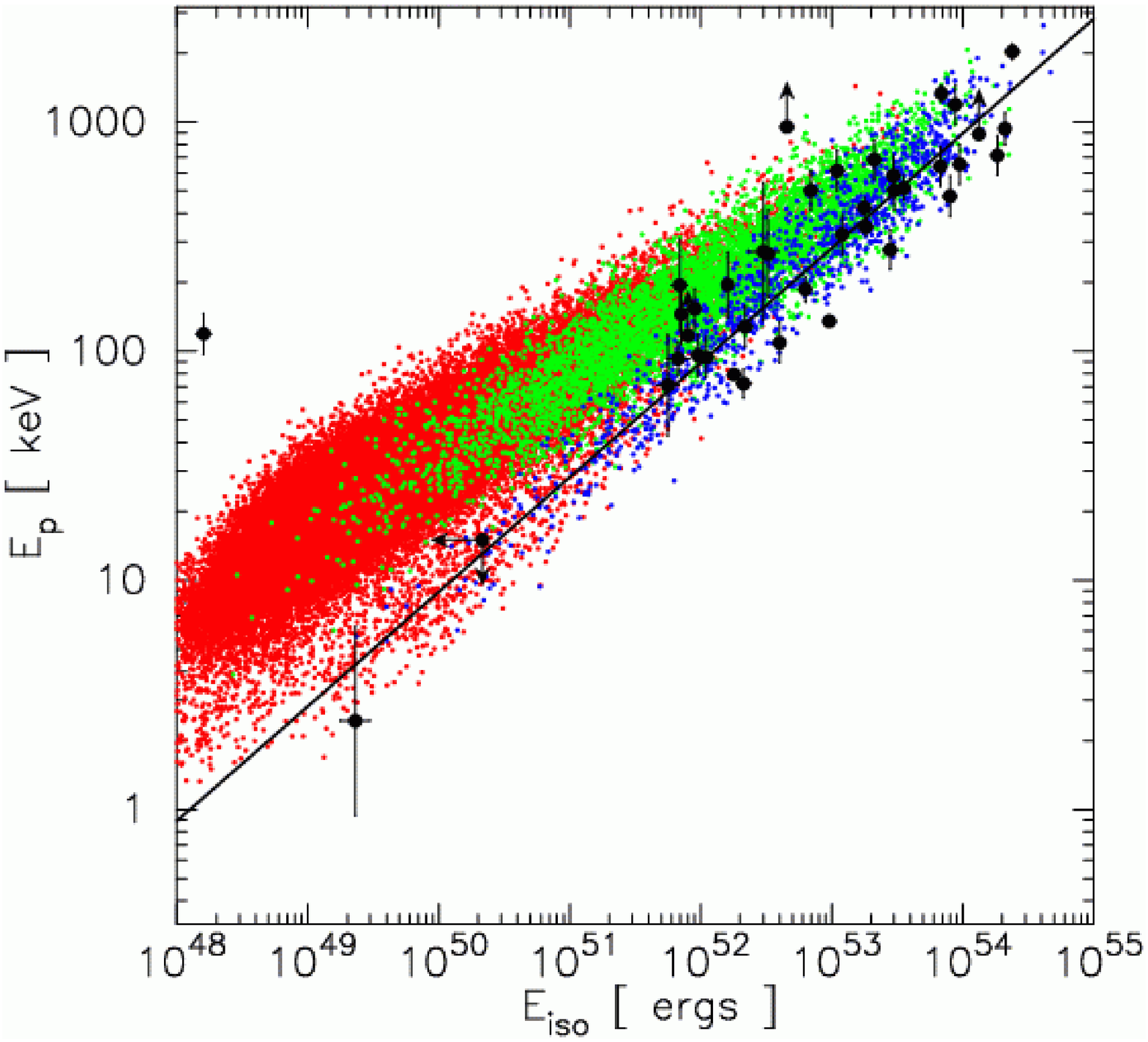}}}
\rotatebox{270}{\resizebox{5.3cm}{!}{\includegraphics{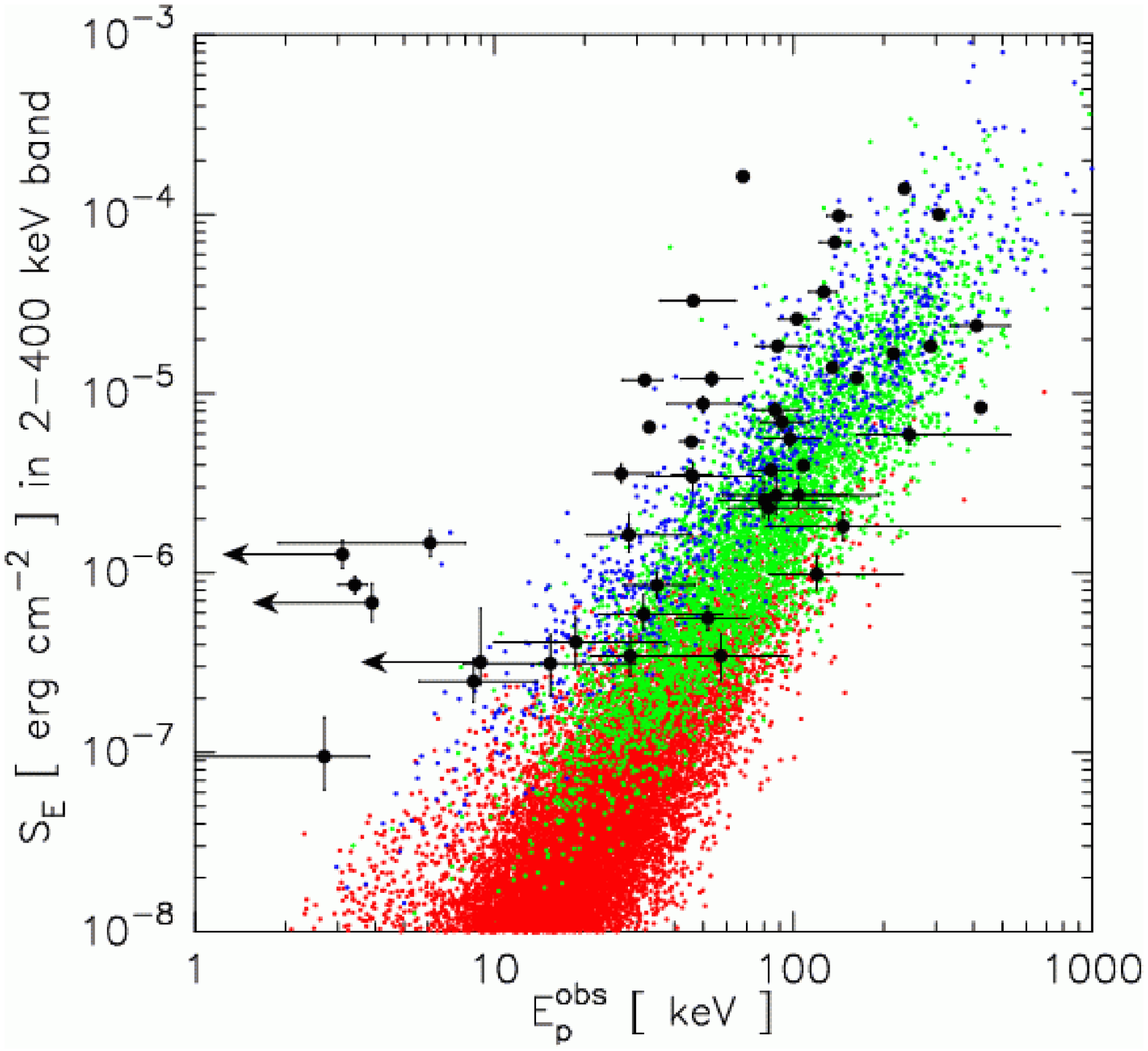}}}
\end{center}
\caption{THVOA1 model, which explains both GRBs and XRFs by a wide
distribution of jet opening-angles, using $\gamma=100$.  Bursts detected
on-jet (blue), off-jet (green) and non-detected bursts (red) are shown
in the [$\thetan$,$\thetav$]-plane (left), [$\eiso$,$\ep$]-plane
(center) and the [$\eop$,$S_{E}$(2-400 keV)]-plane (right), with
observed data from \bsax and \hetetwo overplotted.}
\label{fig_thvoa1}
\end{figure*}

\begin{figure*}[htb]
\begin{center}
\rotatebox{270}{\resizebox{5.3cm}{!}{\includegraphics{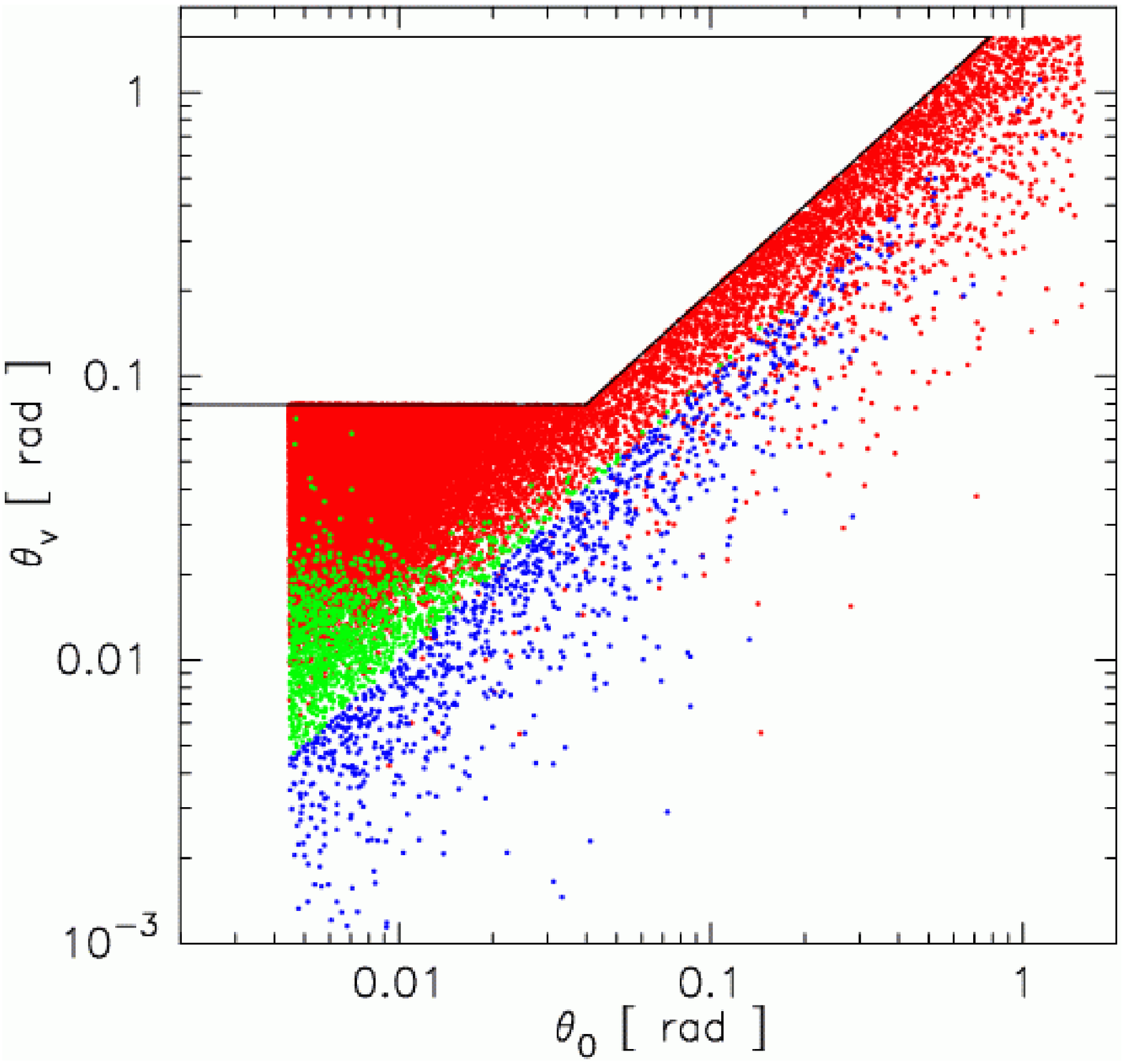}}}
\rotatebox{270}{\resizebox{5.3cm}{!}{\includegraphics{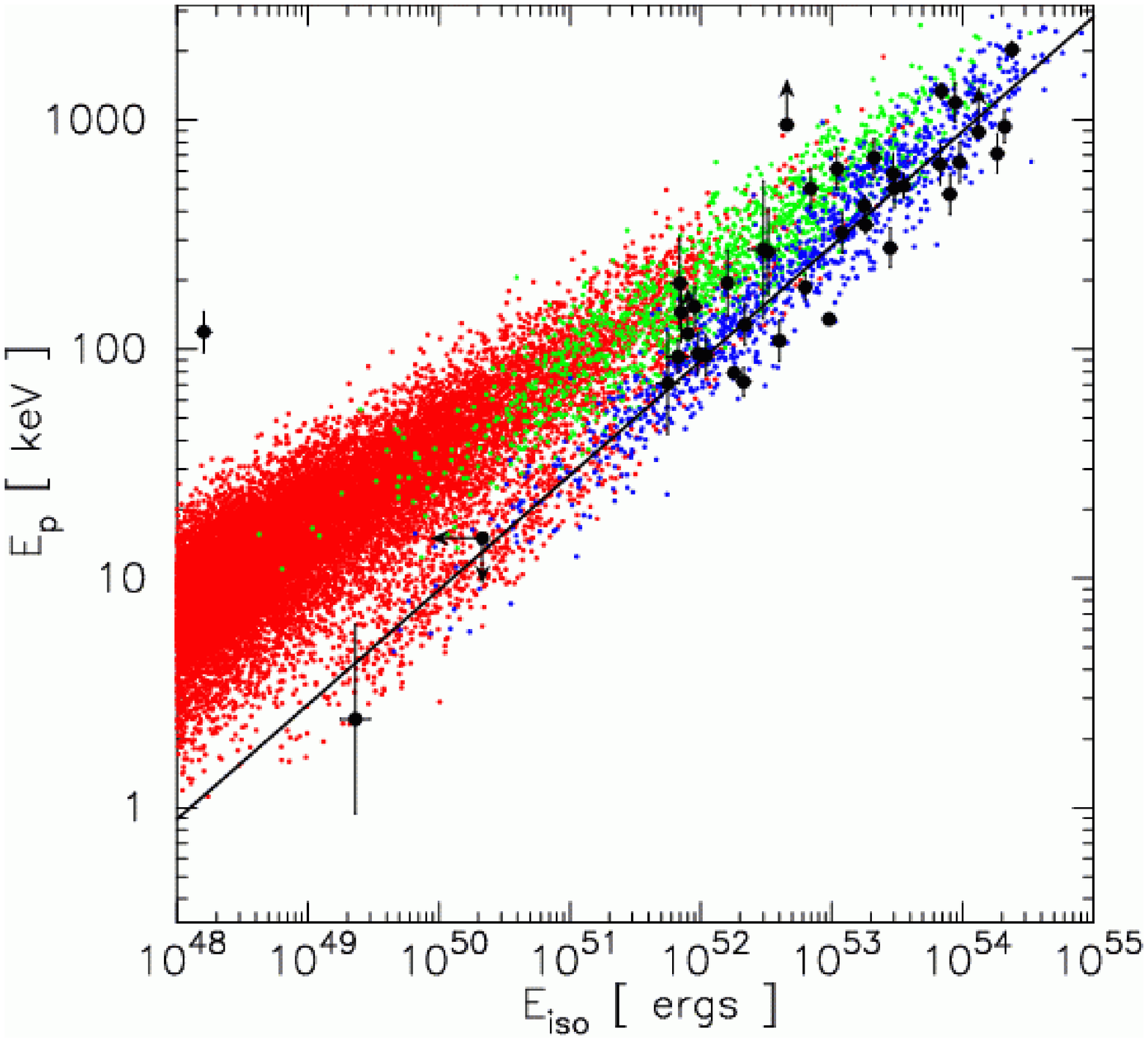}}}
\rotatebox{270}{\resizebox{5.3cm}{!}{\includegraphics{f6c.eps}}}
\end{center}
\caption{THVOA2 model, which explains both GRBs and XRFs by a wide
distribution of jet opening-angles, using $\gamma=300$.  Bursts detected
on-jet (blue), off-jet (green) and non-detected bursts (red) are shown
in the [$\thetan$,$\thetav$]-plane (left), [$\eiso$,$\ep$]-plane
(center) and the [$\eop$,$S_{E}$(2-400 keV)]-plane (right), with
observed data from \bsax and \hetetwo overplotted.}
\label{fig_thvoa2}
\end{figure*}

I draw $\ojet=2\pi(1-\cos\thetan)$ values from a power-law distribution
given by $f_{0}\,d\ojet \propto \ojet^{-2}\,d\ojet$, defined from $2\pi$
to $2\pi \times 10^{-5}$ sr.  $\egt$ is drawn from a narrow lognormal
distribution centered on $1.2 \times 10^{49}$ erg.  The lower central
point for the $\egt$ distribution is a requirement for including in a
unified model those events with measured $\eiso$ values that are smaller
than the usual standard energy of $\sim 10^{51}$ ergs.  I consider
$\gamma=100$ (THVOA1, Figure \ref{fig_thvoa1}) and $\gamma=300$ (THVOA2,
Figure \ref{fig_thvoa2}).

As can be seen from the upper-left panels of Figures \ref{fig_thvoa1}
and \ref{fig_thvoa2}, the relative importance of off-jet events
increases for models with a population of very small opening-angles. 
This is mainly due to the fact that for constant $\egamma$, narrower
jets will have larger $\eiso$ values.  Such bursts will be brighter and
therefore detectable at larger $\thetav$.  Narrow jets are also more
likely to be viewed off-jet than wider jets.  Finally, smaller values
of $\thetan$ give rise to trajectories in the [$\eiso$,$\ep$]-plane that
differ more conspicuously from the $\epei$ relation.  

The large population of very small jet opening angles in these models
also implies a much larger true burst rate than in model Y04.  As Table
\ref{table_stats} shows, for these two models the WXM will detect 1 out
of 2570 and 6470 bursts respectively, as compared to 1 out of 150 bursts
for the Y04 model.\footnote{In calculating the detected fractions, I
correct for the truncation in the simulation parameter space described
in \S 2.  See \S 4 for more details.} 

For larger values of $\gamma$, the emission curves in Figure
\ref{carlo_form} drop off faster away from the edge of the jet. 
Comparing the upper-right panels of Figures \ref{fig_thvoa1} and
\ref{fig_thvoa2} illustrates the fact that larger values of $\gamma$
reduce the percentage of off-jet events observed by the WXM, and
consequently the percentage of bursts seen away from the $\epei$
relation.  This population of events is fairly conspicuous in the
$\gamma=100$ case ($83$\% of detected bursts are off-jet events) and
less so for $\gamma=300$ ($56$\%).

Furthermore, this figure highlights the fact that observed XRFs are more
easily explained by wide opening-angle jets than by off-jet emission. 
No matter the value of $\gamma$, observed XRFs inhabit different regions
of the data planes than do off-jet events.

\subsection{Models Matching the $\epeg$ Relation}

Reproducing the $\epeg$ relation \citep{ghirlanda2004} is an important
test of any jet model.  Discovery of this relation post-dated
\cite{ldg2005}, and so was not addressed in that paper.  Here I
construct models that satisfy this relation, in addition to the
constraints of the earlier models (for more details see
\cite{dlg2005c}).

The $\epei$ and $\epeg$ relations can be mutually satisfied in the
top-hat variable opening-angle jet model by imposing a relation between
$\ojet$ and $\egamma$.  The following expressions hold only for the
on-jet events; the off-jet events behave differently.  Combining
Equations \ref{eqn_amati} and \ref{eqn_ggl} with the definition
\beq
\eiso = \frac{\egamma}{\ojet/2\pi} = \frac{\egamma}{1-\cos\thetan},
\eeq
I find
\beq
\ep = C_{A} \left( \frac{\egamma \cdot 2\pi}{E_{A} \cdot \ojet} 
	\right)^{\alpha}
	= C_{G} \left( \frac{\egamma}{E_{G}} \right)^{\beta}
\eeq
and solving for $\ojet$ gives the following expression
\beq
\ojet = 2\pi (1-\cos\thetan)
	= 2\pi \left( \frac{C_{A}}{C_{G}} \cdot 
	\frac{E_{G}^{\beta}}{E_{A}^{\alpha}} \right)^{1/\alpha}
	\egamma^{(\alpha-\beta)/\alpha}.
\label{eqn_ojet}\eeq

\begin{figure*}[htb]
\begin{center}
\rotatebox{270}{\resizebox{5.3cm}{!}{\includegraphics{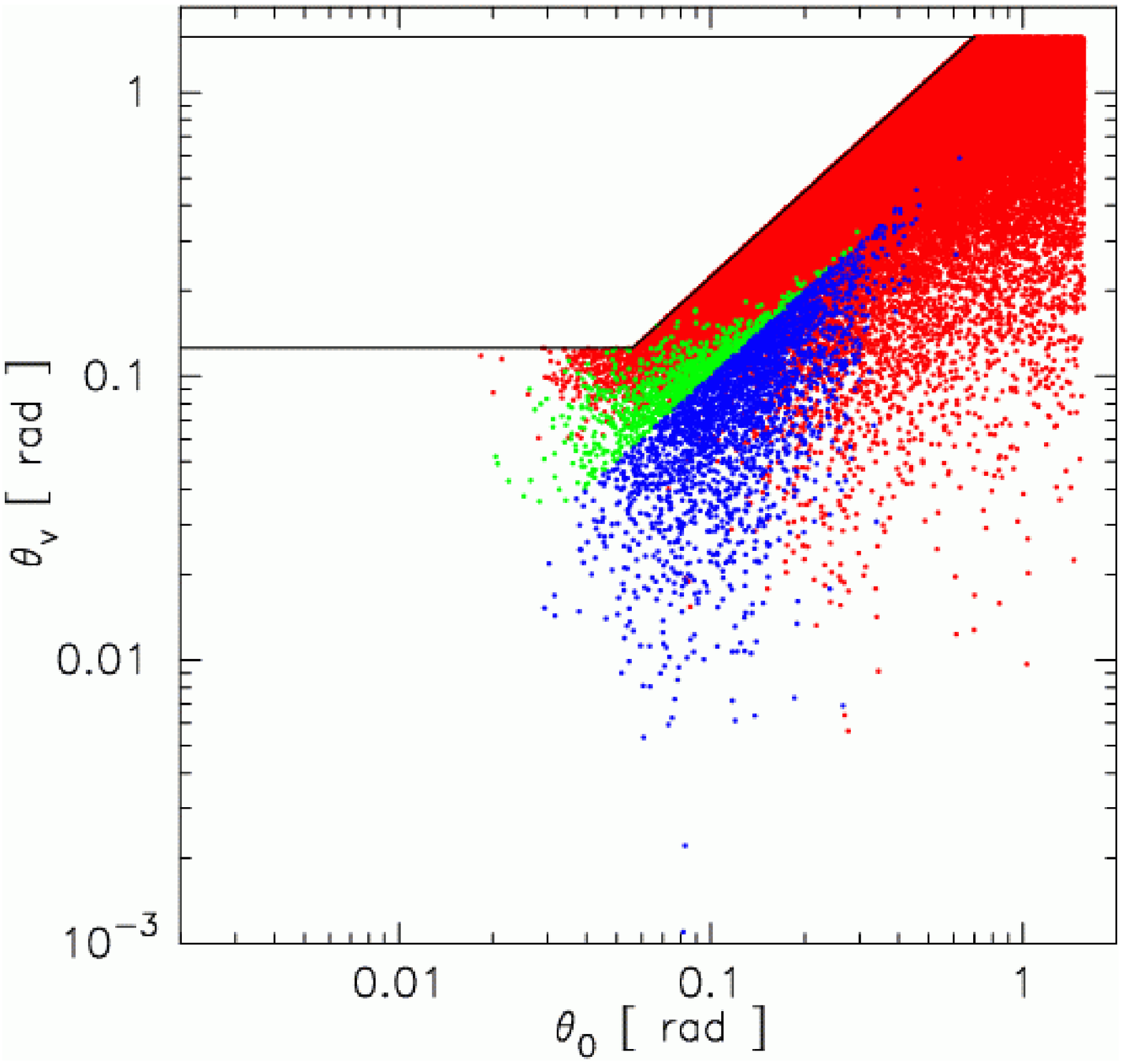}}}
\rotatebox{270}{\resizebox{5.3cm}{!}{\includegraphics{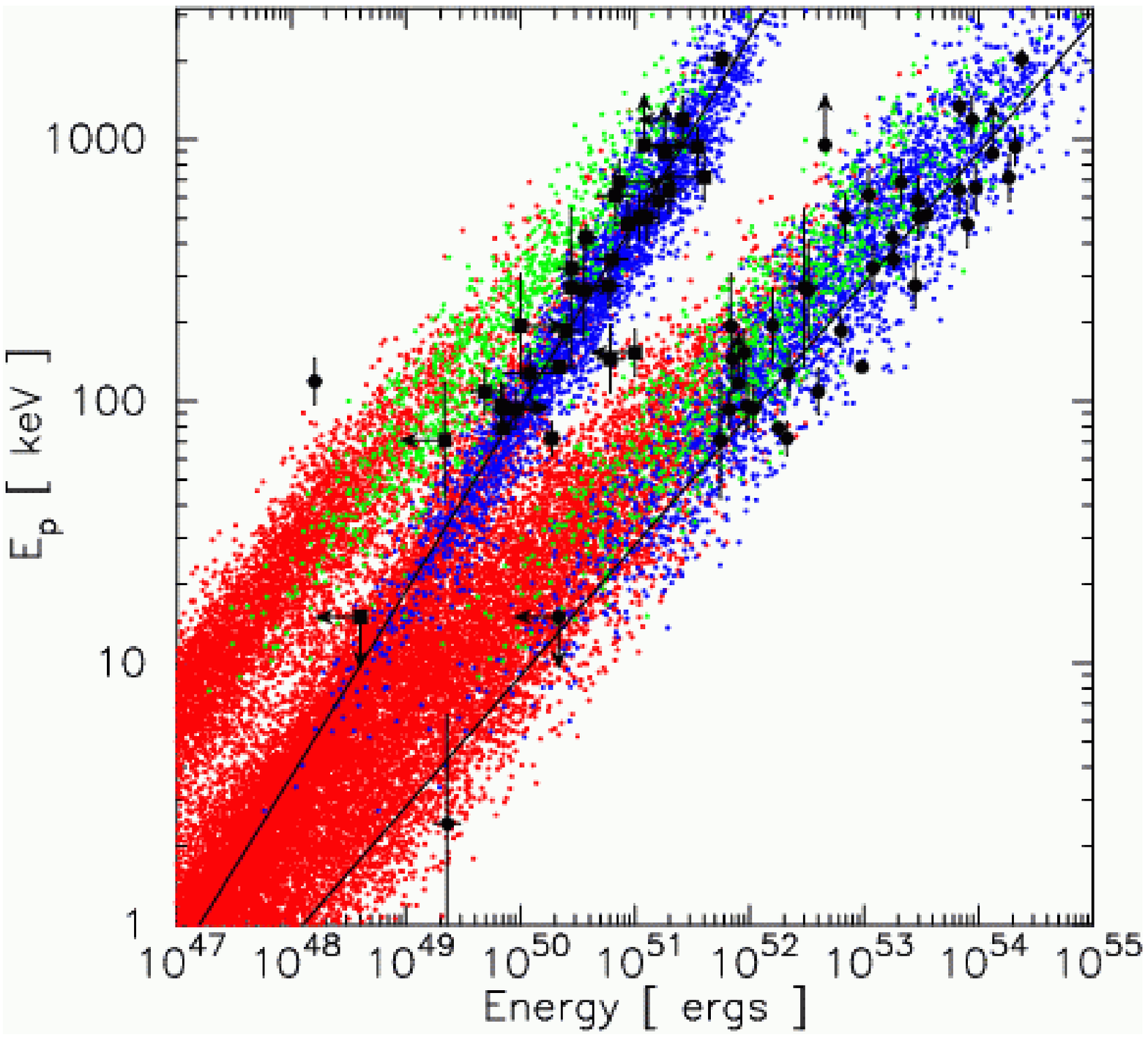}}}
\rotatebox{270}{\resizebox{5.3cm}{!}{\includegraphics{f7c.eps}}}
\end{center}
\caption{GGL1 model, which picks parameters to match both the $\epei$
and $\epeg$ relations, using $\gamma=100$.  Bursts detected on-jet
(blue), off-jet (green) and non-detected bursts (red) are shown in the
[$\thetan$,$\thetav$]-plane (left), [$E$,$\ep$]-plane (center) and the
[$\eop$,$S_{E}$(2-400 keV)]-plane (right), with observed data from \bsax
and \hetetwo overplotted.}
\label{fig_ggl1}
\end{figure*}

\begin{figure*}[htb]
\begin{center}
\rotatebox{270}{\resizebox{5.3cm}{!}{\includegraphics{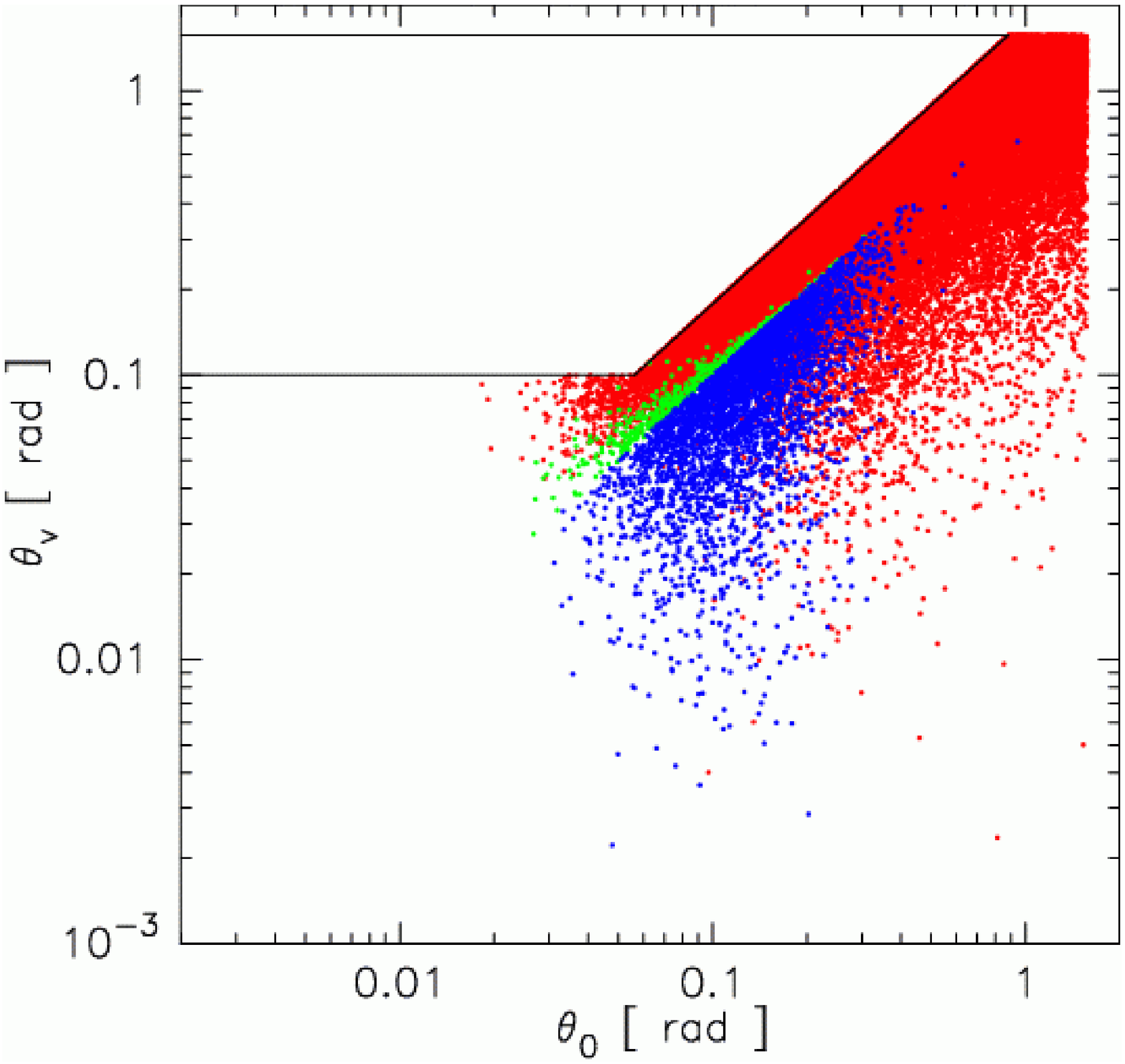}}}
\rotatebox{270}{\resizebox{5.3cm}{!}{\includegraphics{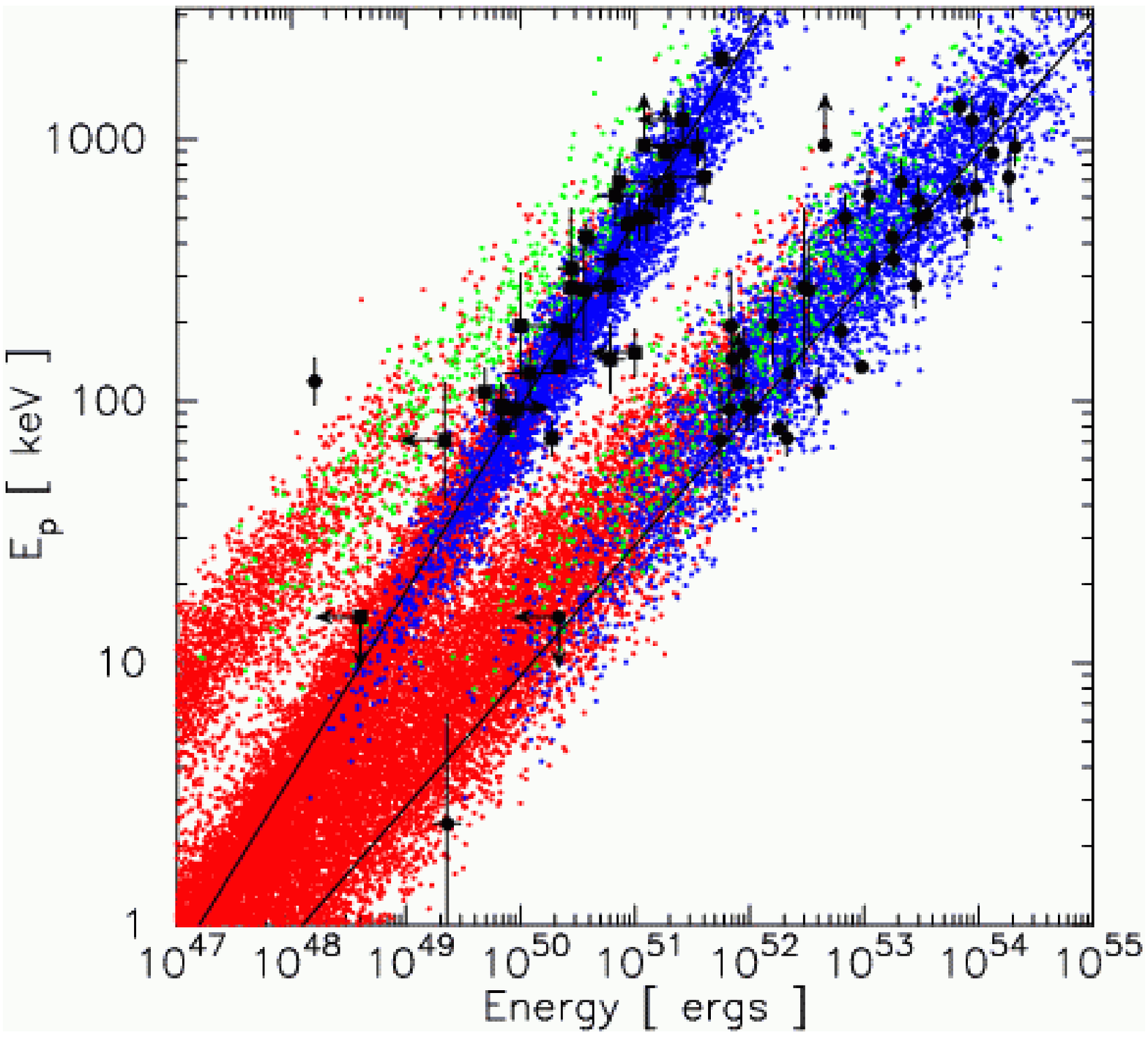}}}
\rotatebox{270}{\resizebox{5.3cm}{!}{\includegraphics{f8c.eps}}}
\end{center}
\caption{GGL2 model, which picks parameters to match both the $\epei$
and $\epeg$ relations, using $\gamma=300$.  Bursts detected on-jet
(blue), off-jet (green) and non-detected bursts (red) are shown in the
[$\thetan$,$\thetav$]-plane (left), [$E$,$\ep$]-plane (center) and the
[$\eop$,$S_{E}$(2-400 keV)]-plane (right), with observed data from \bsax
and \hetetwo overplotted.}
\label{fig_ggl2}
\end{figure*}

I consider two different models in which $\ojet$ and $\egt$ are
specified by imposing this relationship. The natural minimum value for
the distribution of $\egt$ is the point $E^{*}$ at which $\egamma =
\eiso$, which is found to be 
\beq 
E^{*} = \left( \frac{C_{A} E_{G}^{\beta}}{C_{G} E_{A}^{\alpha}} 
	\right)^{1/(\beta-\alpha)}. 
\eeq
Current parameters for the correlations give a value of $E^{*} = 3
\times 10^{44}$ ergs, which is well below current observational
thresholds.  Values of $\egt$ are generated by drawing from a power-law
distribution that gives equal numbers of bursts per decade, and is
defined from $E^{*}$ through a maximum value ($\egt = 3.16 \times
10^{51}$ ergs) that encompasses the largest observed $\egamma$ value
($\egi = 5.75 \times 10^{51}$ ergs for GRB 990123).  To avoid a sharp
cutoff at the minimum and maximum values of $\egamma$, I include an
additional smearing function in the simulated value of $\egt$ with a
lognormal width of $0.3$ decades.  $\ojet$ (and hence $\thetan$) are
found via Equation \ref{eqn_ojet}, and the rest of the simulations
proceed as above.  Results for $\gamma=100$ (GGL1) and $\gamma=300$
(GGL2) are shown in Figures \ref{fig_ggl1} and \ref{fig_ggl2},
respectively.

Current data seems to indicate that the $\epeg$ relation has a narrower
distribution about the best-fit line than does the $\epei$ relation
\citep{ghirlanda2004}.  Adding a Gaussian smearing function to $C_{A}$,
as was done in the above models, produces equal widths for both
distributions.  To broaden the $\epei$ relation, I introduce an
additional Gaussian smearing function into the relation between
$\egamma$ and $\ojet$.  Equation \ref{eqn_ojet} then becomes
\beq
\ojet = \frac{2\pi}{C_{\Omega}} \left( \frac{C_{A}}{C_{G}} \cdot 
	\frac{E_{G}^{\beta}}{E_{A}^{\alpha}} \right)^{1/\alpha}
	\egamma^{(\alpha-\beta)/\alpha},
\eeq
where $C_{\Omega}$ is centered on $1.0$ and has a lognormal width of
$0.3$.  Combined with the above value of the smearing in $C_{A}$, this
approach and value of $C_{\Omega}$ gives good agreement with both
observed distribution widths.

An immediate consequence of the $\ojet$-$\egamma$ correlation is a lack
of very small jet opening-angles and a narrower range of $\thetan$
(compare the upper-right panels of Figures \ref{fig_thvoa2} and
\ref{fig_ggl2}).  The bursts observed off-jet lie closer to the $\epei$
relation than in models without the correlation (compare the upper-left
panels of Figures \ref{fig_thvoa2} and \ref{fig_ggl2}).  This is due to
the narrower range of $\thetan$ values mentioned above; larger $\thetan$
values produce trajectories in the [$\eiso$,$\ep$]-plane that closer
approximate a $0.5$ power-law.  The off-jet bursts also deviate further
from the $\epeg$ relation than from the $\epei$ relation.  This is due
to that fact that, given an on-jet and an off-jet event that lie near
each other on the $\epei$ relation, the off-jet event is more likely to
have a smaller $\thetabr$ than the on-jet event, thereby giving a
smaller $\egi$ value.  

In comparison with the two THVOA models, these models exhibit much
smaller true burst rates (1 detection for every 165 and 197 bursts,
respectively) and smaller fractions of off-jet events ($26$\% and $10$\%
of detected bursts, respectively).  The additional Gaussian smearing
function, $C_{\Omega}$, has the effect of blurring the clear separation
of blue and green points along the $\epei$ relation (compare THVOA1 and
GGL1), but not along the $\epeg$ relation in which $C_{\Omega}$ plays no
role.

\subsection{Models with $\ojet$-$\gamma$ Correlations}

Finally, I investigate the effect of an additional correlation between
$\ojet$ and $\gamma$.  If narrower jets have larger bulk $\gamma$
values, this could reduce the importance of off-jet emission even
further.  The exact relationship between the bulk $\gamma$ of the
material and the opening-angle of the jet is unknown.  Extending the GGL
models in the previous section, I consider a simple model (GCOR) in
which $\gamma$ is given by $\gamma \propto \ojet^{-1}$.  I fix the
normalization by setting $\gamma = 1000$ for the bursts with the
narrowest jets.

\begin{figure*}[htb]
\begin{center}
\rotatebox{270}{\resizebox{5.3cm}{!}{\includegraphics{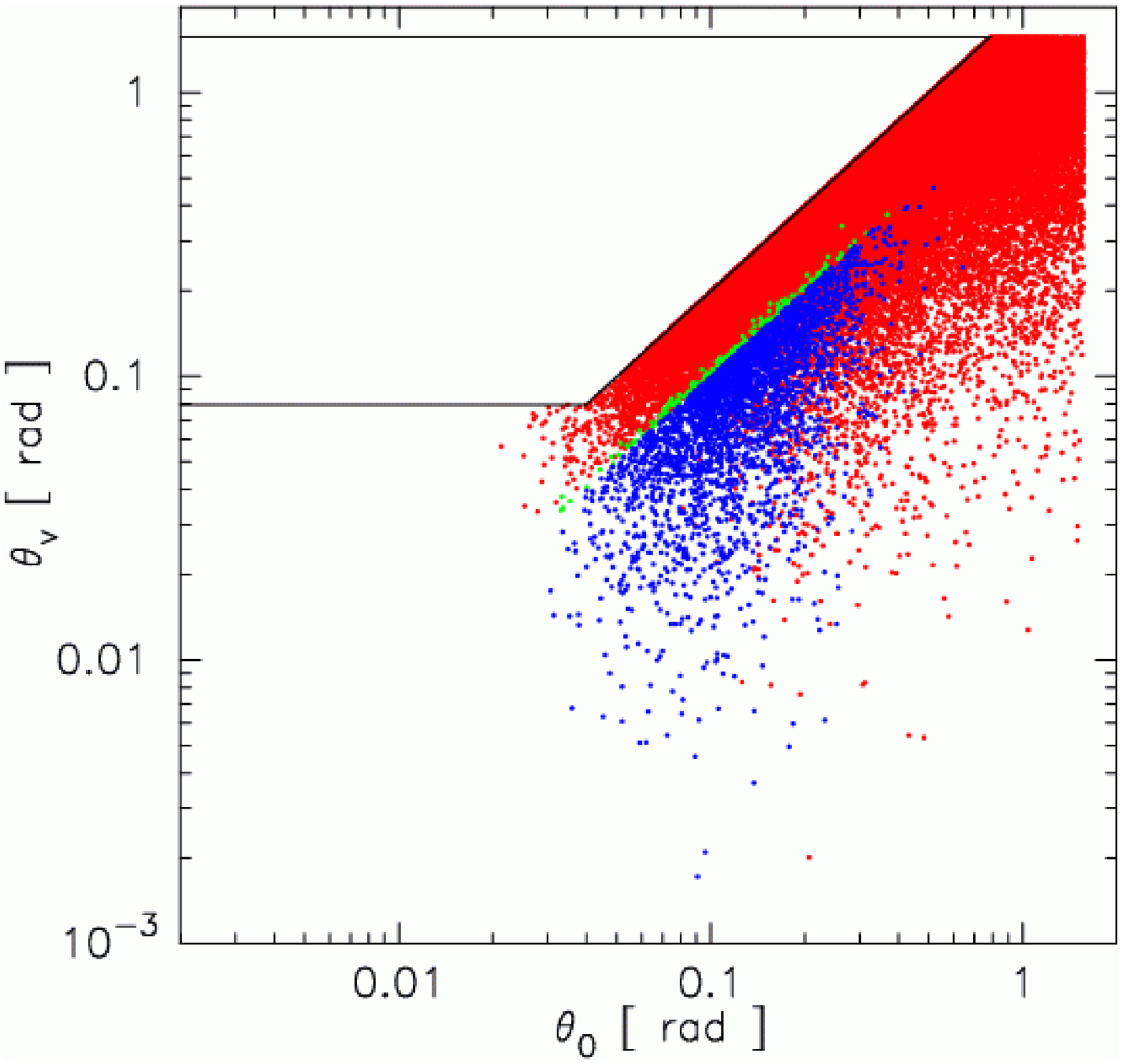}}}
\rotatebox{270}{\resizebox{5.3cm}{!}{\includegraphics{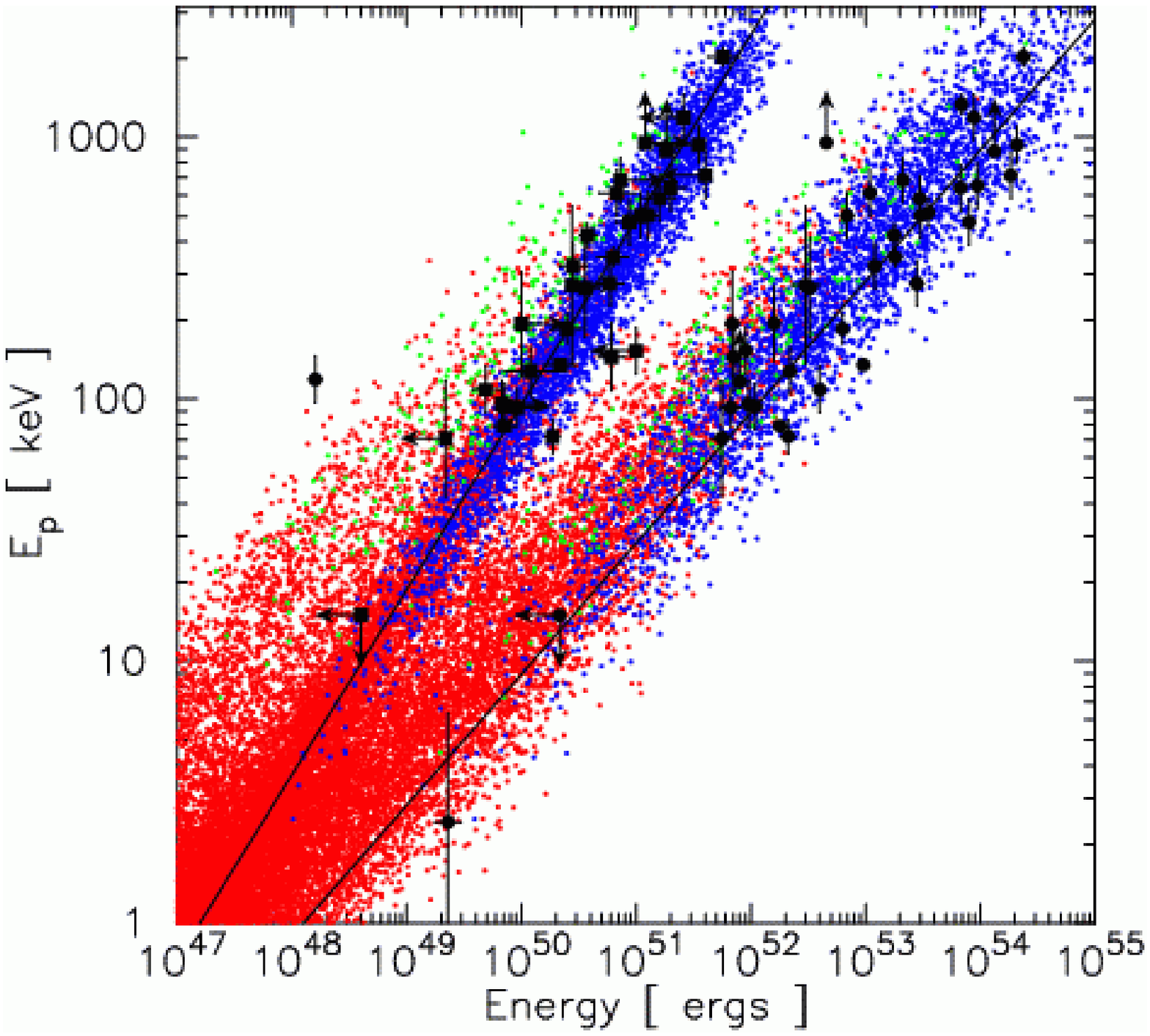}}}
\rotatebox{270}{\resizebox{5.3cm}{!}{\includegraphics{f9c.eps}}}
\end{center}
\caption{Model GCOR where $\gamma \propto \ojet^{-1}$ and $\gamma=1000$
at the minimum value of $\ojet$.  Bursts detected on-jet (blue), off-jet
(green) and non-detected bursts (red) are shown in the
[$\thetan$,$\thetav$]-plane (left), [$E$,$\ep$]-plane (center) and the
[$\eop$,$S_{E}$(2-400 keV)]-plane (right), with observed data from \bsax
and \hetetwo overplotted.}
\label{fig_gamma}
\end{figure*}

Figure \ref{fig_gamma} shows that imposing this correlation greatly
reduces the percentage of bursts seen off-jet ($<7$\% of detected bursts
are seen off-jet).  Physically this is due to a combination of several
effects.  It is more probable to observe narrow jets at viewing angles
slightly outside the jet than inside the jet, yet for such narrow jets a
large value of $\gamma$ ensures that the detectability of such
slightly-off-jet bursts drops off very quickly with viewing angle. 
Broader jets may have smaller values of $\gamma$, but they are less
likely to be observed off-jet.

\section{Discussion}
\label{cha:discuss}

Table \ref{table_stats} summarizes the detected fractions and off-jet
fractions for the six models.  The detected fraction has direct
implications for the total rate of GRBs.  The ratio of the observed rate
of Type Ic supernovae to the observed rate of GRBs is roughly $10^{5}$
\citep{lamb1999,lamb2000}, and therefore the ratio of the observed rate
of Type Ic supernovae to the true rate of GRBs is that value times the
detected fraction for the model.  Due to their very narrow jets, the
THVOA models have the smallest detected fractions of the six models
presented here.  For the model THVOA2, GRBs may comprise an appreciable
fraction of all observed Type Ic supernovae, but for all other models,
the true rate of GRBs is much smaller than the observed supernova rate. 

\begin{table}[htb]
\begin{center}
\begin{tabular}{clc}
\hline\hline
Model & Detected Fraction & Off-Jet Fraction \\
\hline
Y04    & $0.00659 = 1/151.7$  & $0.339$ \\  
THVOA1 & $0.00039 = 1/2571.1$ & $0.829$ \\  
THVOA2 & $0.00015 = 1/6467.3$ & $0.561$ \\  
GGL1   & $0.00606 = 1/165.0$  & $0.262$ \\  
GGL2   & $0.00507 = 1/197.3$  & $0.103$ \\  
GCOR   & $0.00489 = 1/204.4$  & $0.066$ \\  
\hline\hline
\end{tabular}
\end{center}
\caption{Detection statistics for the six models presented in this
paper.  The second column shows the true rate of detection for all
bursts in the model.  Note that this is not the same as the percentage
of bursts detected in the samples of $50,000$ due to the truncation
described above.  I correct for that truncation in calculating the
detection fraction.  The third column shows the fraction of all detected
bursts that are seen off-jet.}
\label{table_stats}
\end{table}

In the figures above, I compare theoretical models employing the WXM
detector threshold with data compiled from many instruments with varying
detector sensitivities.  To assess the effect of the detector thresholds
in the simulations, I compare the predicted distributions of observed
bursts for the two most successful models (GGL2 and GCOR) using the WXM
instrument and using the GRBM instrument on \bsaxnosp.  For these two
models, I compare the predicted distribution of observed bursts
employing a given instrument threshold only with the data from that
instrument.

\begin{figure*}[htb]
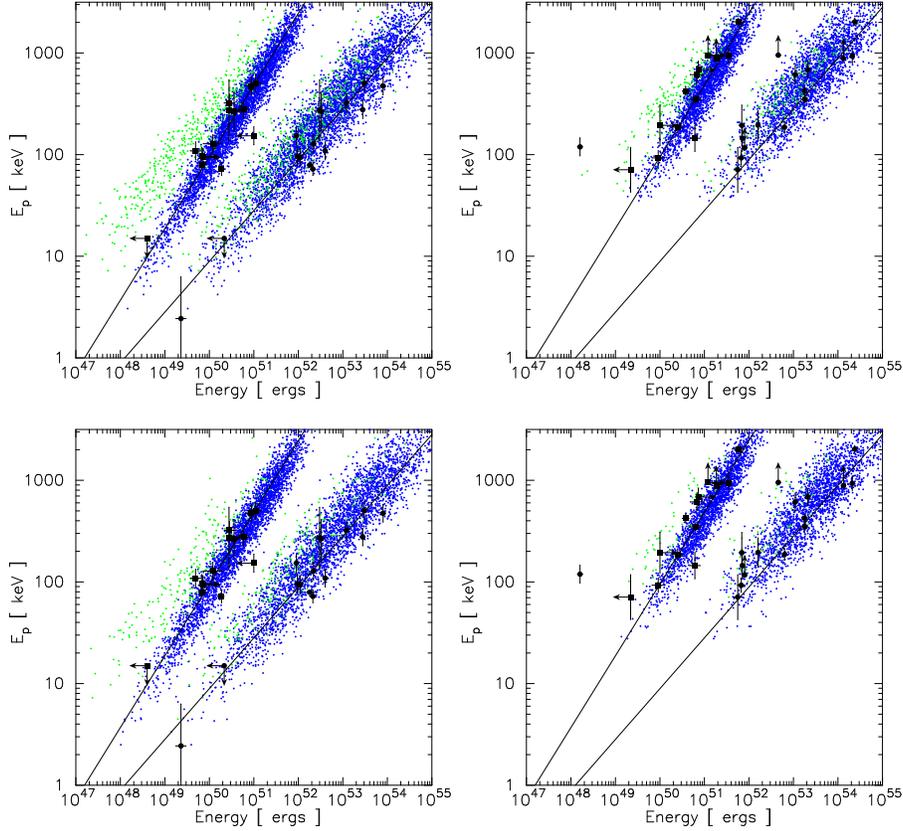

\begin{center}
\rotatebox{270}{\resizebox{5.3cm}{!}{\includegraphics{f10a.eps}}}
\rotatebox{270}{\resizebox{5.3cm}{!}{\includegraphics{f10b.eps}}}
\end{center}
\begin{center}
\rotatebox{270}{\resizebox{5.3cm}{!}{\includegraphics{f10c.eps}}}
\rotatebox{270}{\resizebox{5.3cm}{!}{\includegraphics{f10d.eps}}}
\end{center}
\caption{The effect of different detector thresholds for two models,
GGL2 (upper row) and GCOR (lower row).  Bursts are shown in the
[$E$,$\ep$]-plane, comparing models employing the WXM threshold against
WXM data points (left column) and models employing the \bsax threshold
against \bsax data points (right column).}
\label{hete_bsax_comp}
\end{figure*}

Figure \ref{hete_bsax_comp} shows that the higher triggering threshold
for the GRBM instrument on \bsax prevented that mission from promptly
localizing the fainter, low-$\ep$ XRFs.  In contrast, rapid \hetetwo
localizations of XRFs have provided evidence that the $\epei$ relation
extends to lower $\ep$ values, but \hetetwo has detected fewer
high-$\ep$ bursts than \bsaxnosp.  Therefore, the models presented in
this paper employing the WXM threshold are a good match for the
full-range of observed GRB characteristics.

\begin{figure*}[htb]
\begin{center}
\rotatebox{270}{\resizebox{5.3cm}{!}{\includegraphics{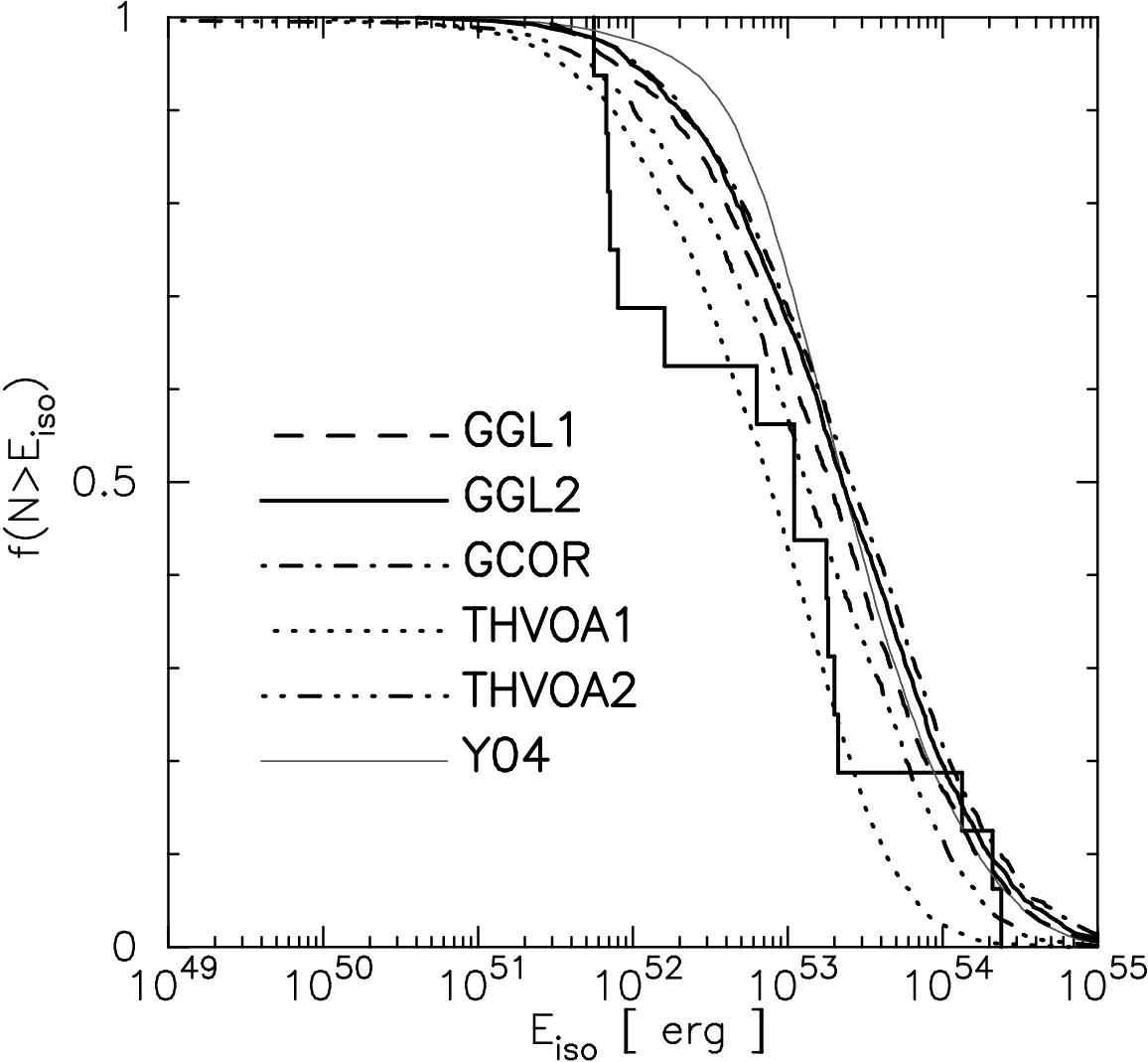}}}
\rotatebox{270}{\resizebox{5.3cm}{!}{\includegraphics{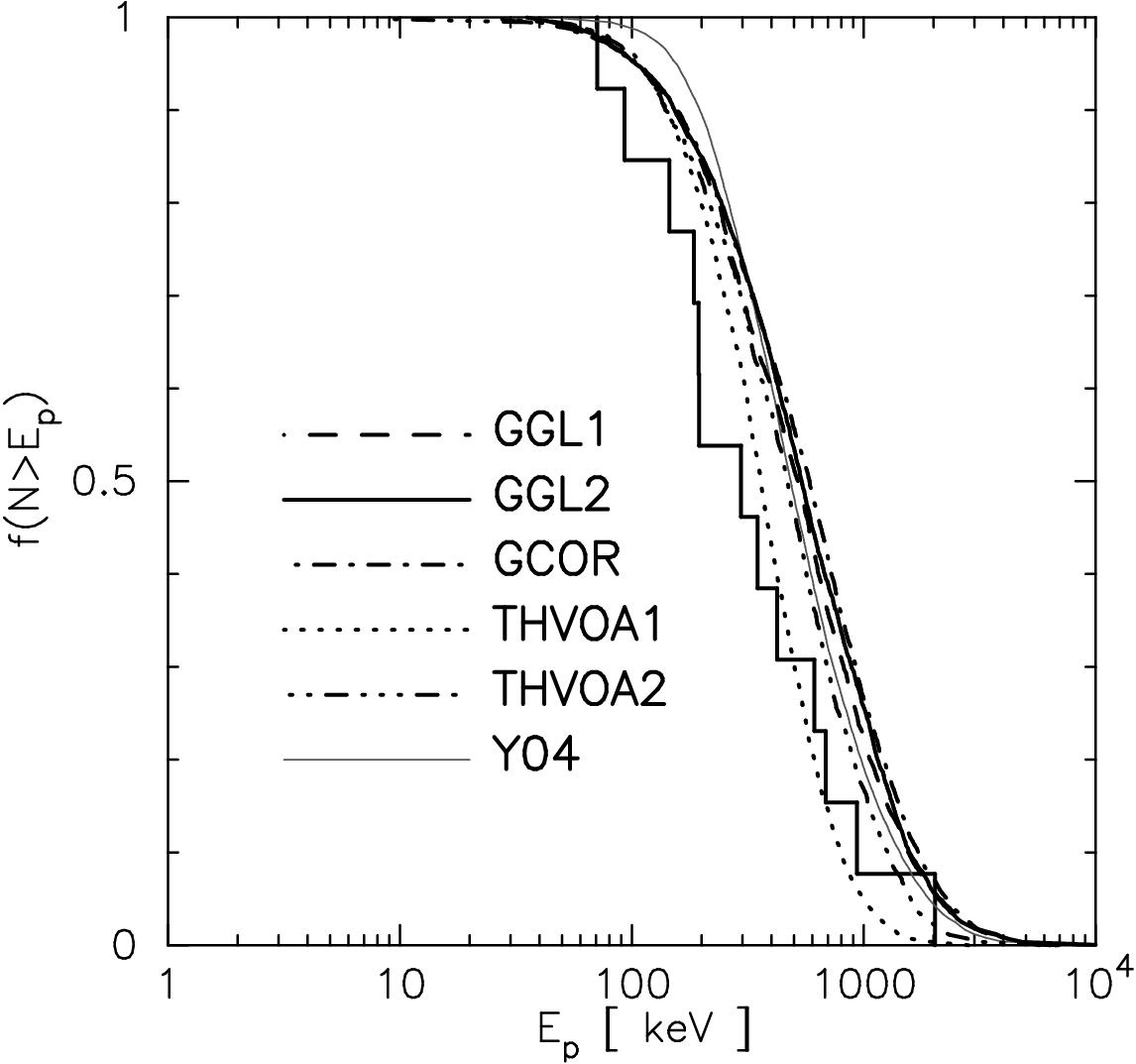}}}
\end{center}
\begin{center}
\rotatebox{270}{\resizebox{5.3cm}{!}{\includegraphics{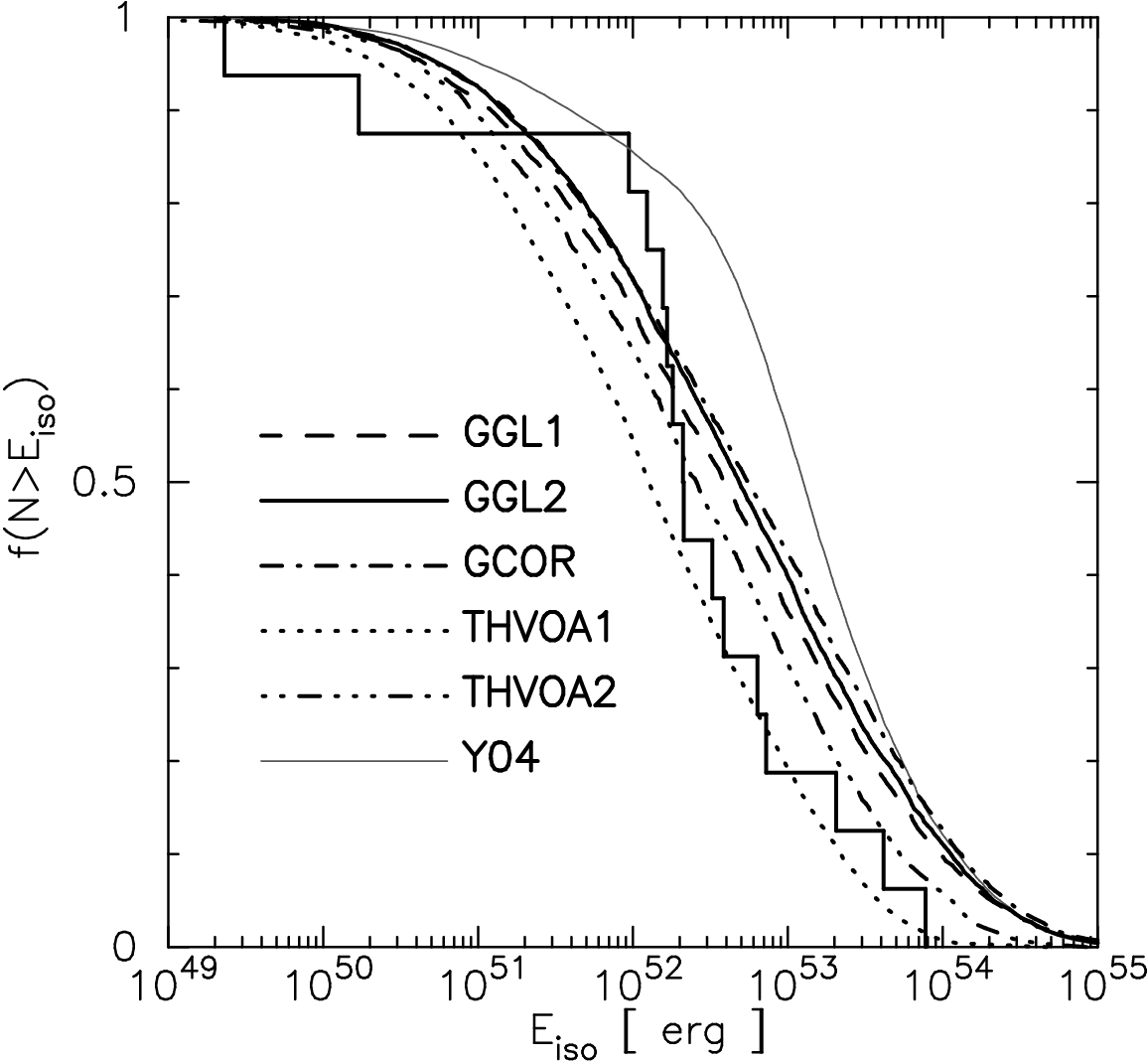}}}
\rotatebox{270}{\resizebox{5.3cm}{!}{\includegraphics{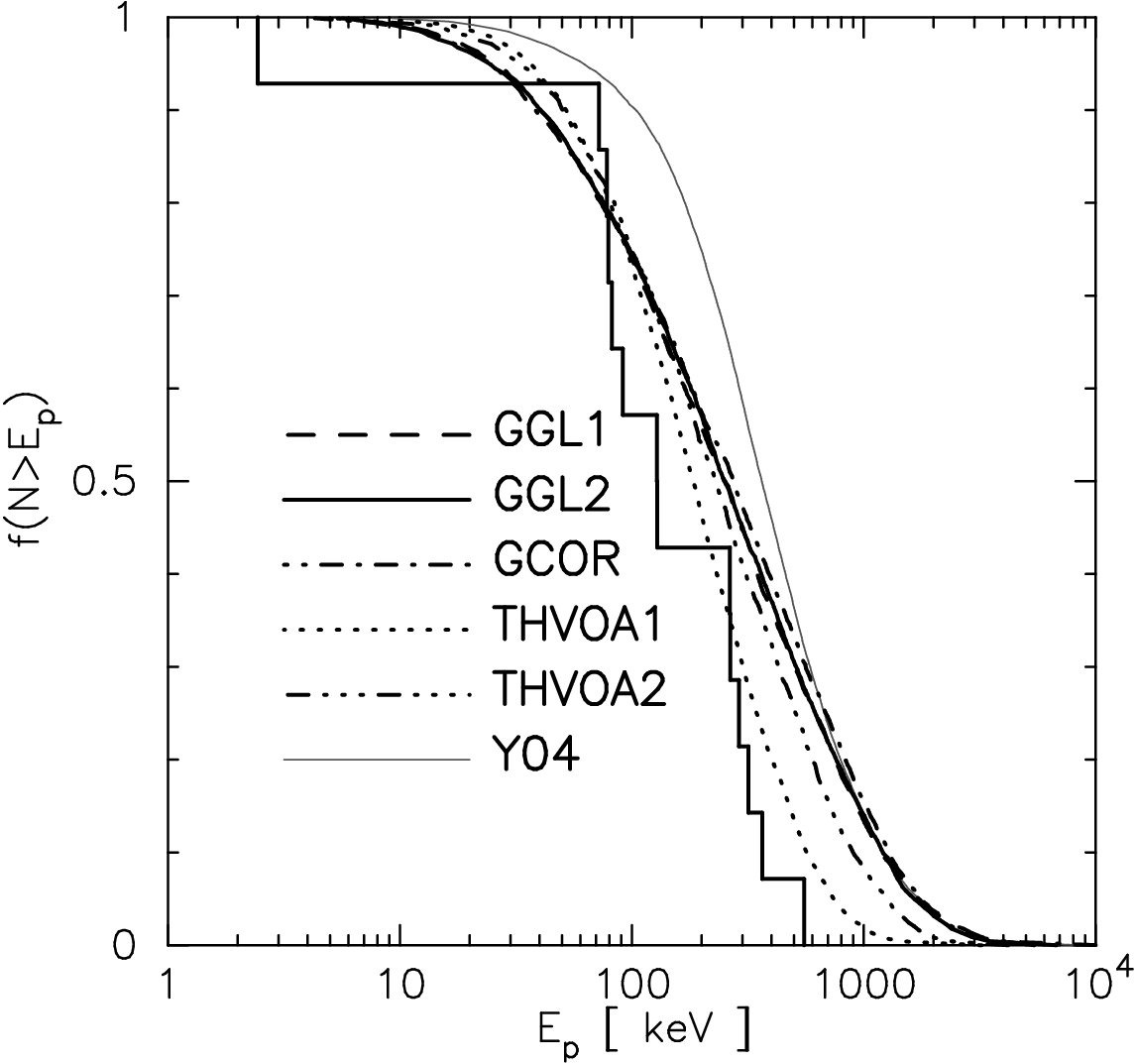}}}
\end{center}
\caption{The cumulative distribution in $\eiso$ (left column) and $\ep$
(right column) for detected bursts in the six models explored in this
paper.  The upper row compares models employing the GRBM detector with
data points observed by \bsaxnosp.  The lower row compares models
employing the WXM detector with data points observed by \hetetwonosp. 
The data used for the $\ep$ and $\eiso$ distributions are the same as in
Figure \ref{datasets}b, broken down by detector.}
\label{cuml_figs_Eiso_Ep}
\end{figure*}

\begin{figure*}[htb]
\begin{center}
\rotatebox{270}{\resizebox{5.3cm}{!}{\includegraphics{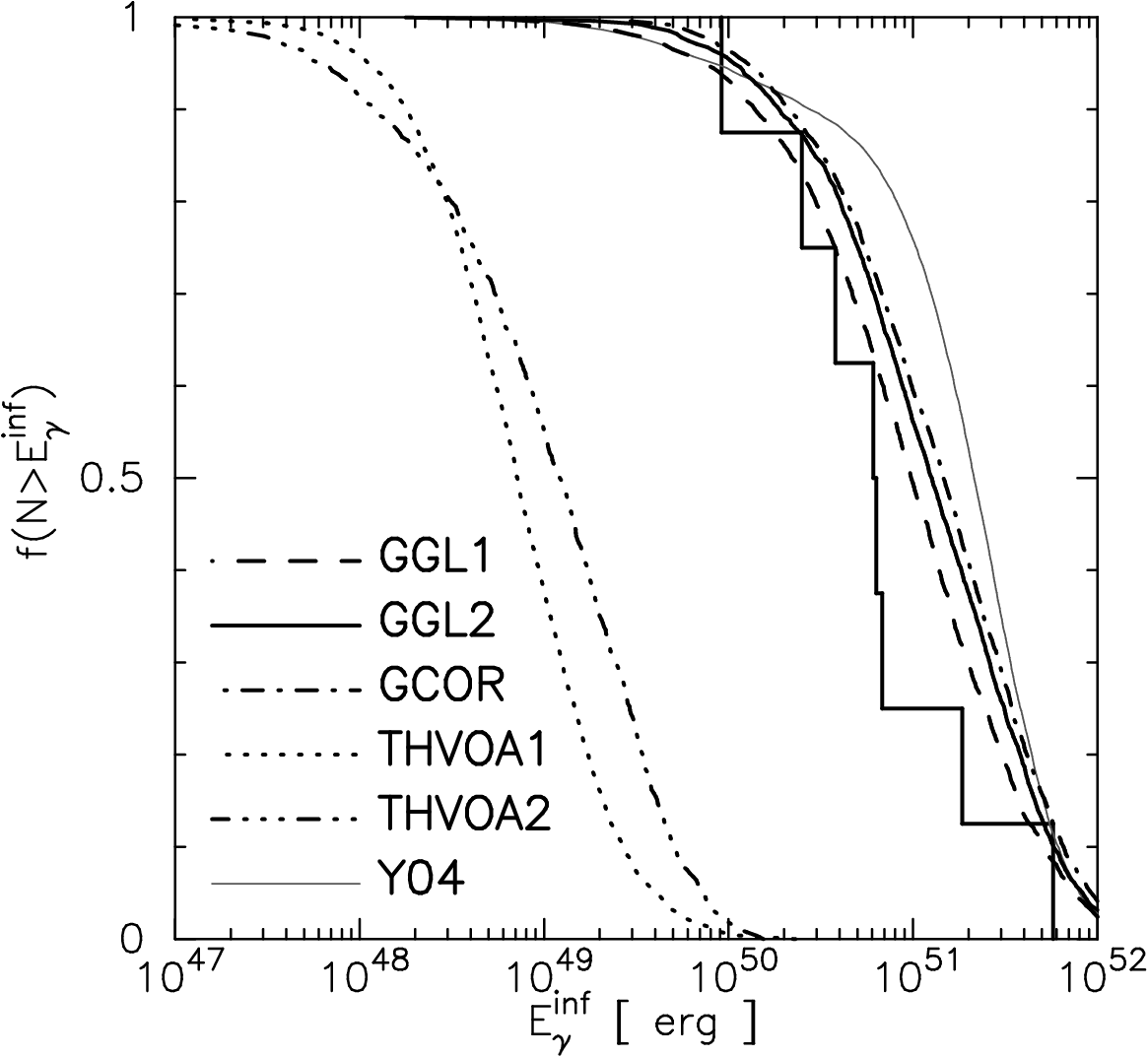}}}
\rotatebox{270}{\resizebox{5.3cm}{!}{\includegraphics{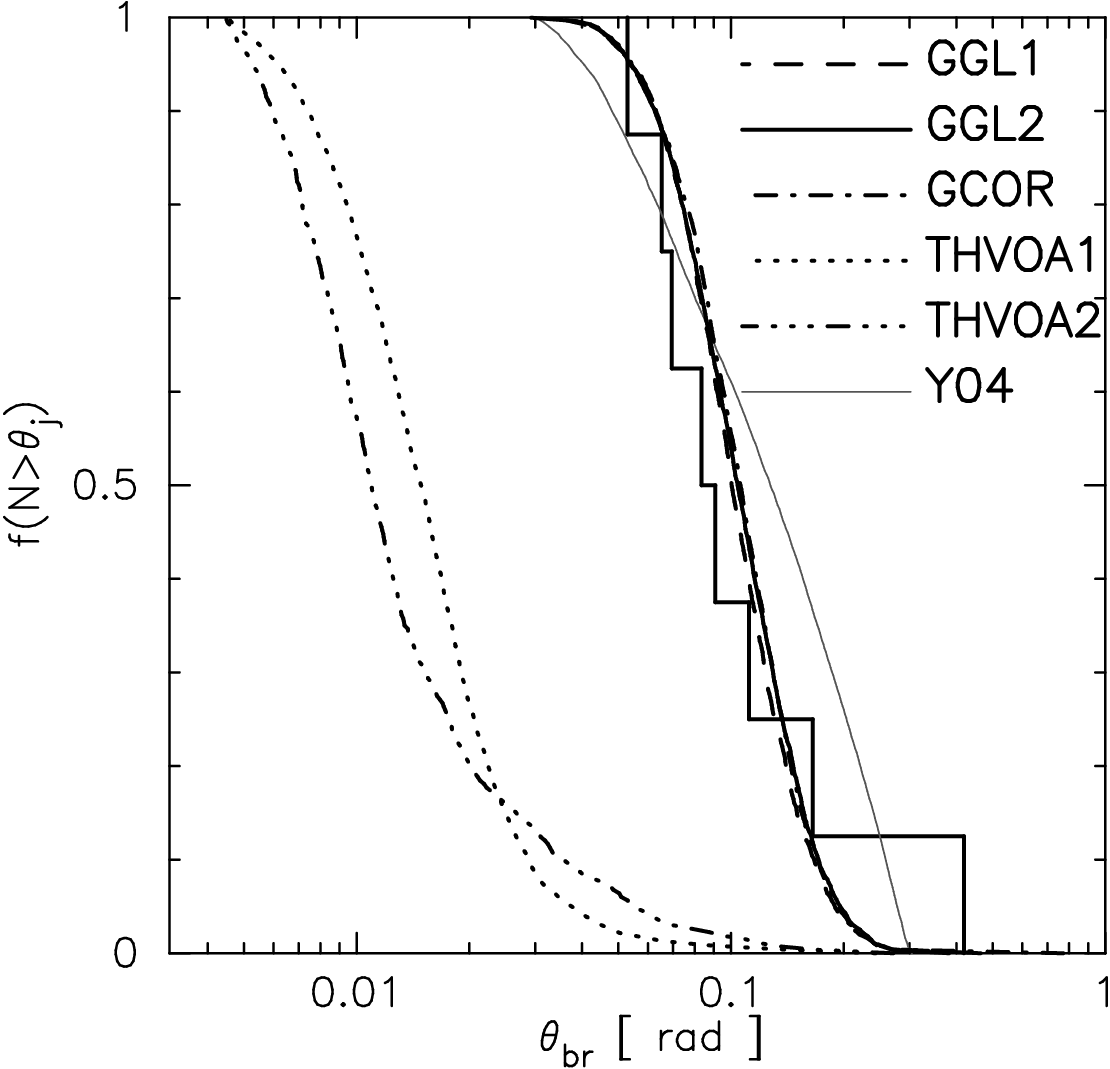}}}
\end{center}
\begin{center}
\rotatebox{270}{\resizebox{5.3cm}{!}{\includegraphics{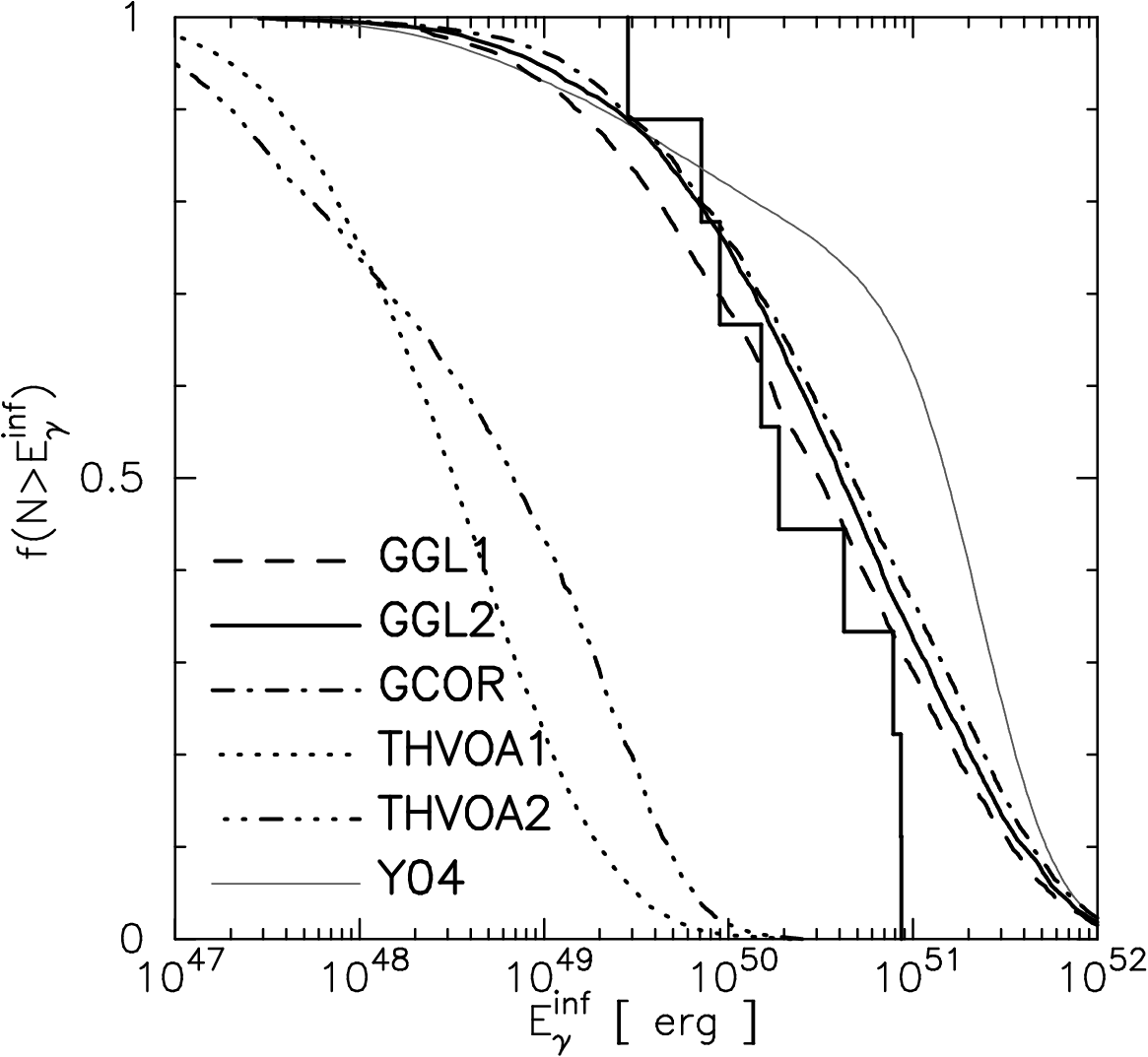}}}
\rotatebox{270}{\resizebox{5.3cm}{!}{\includegraphics{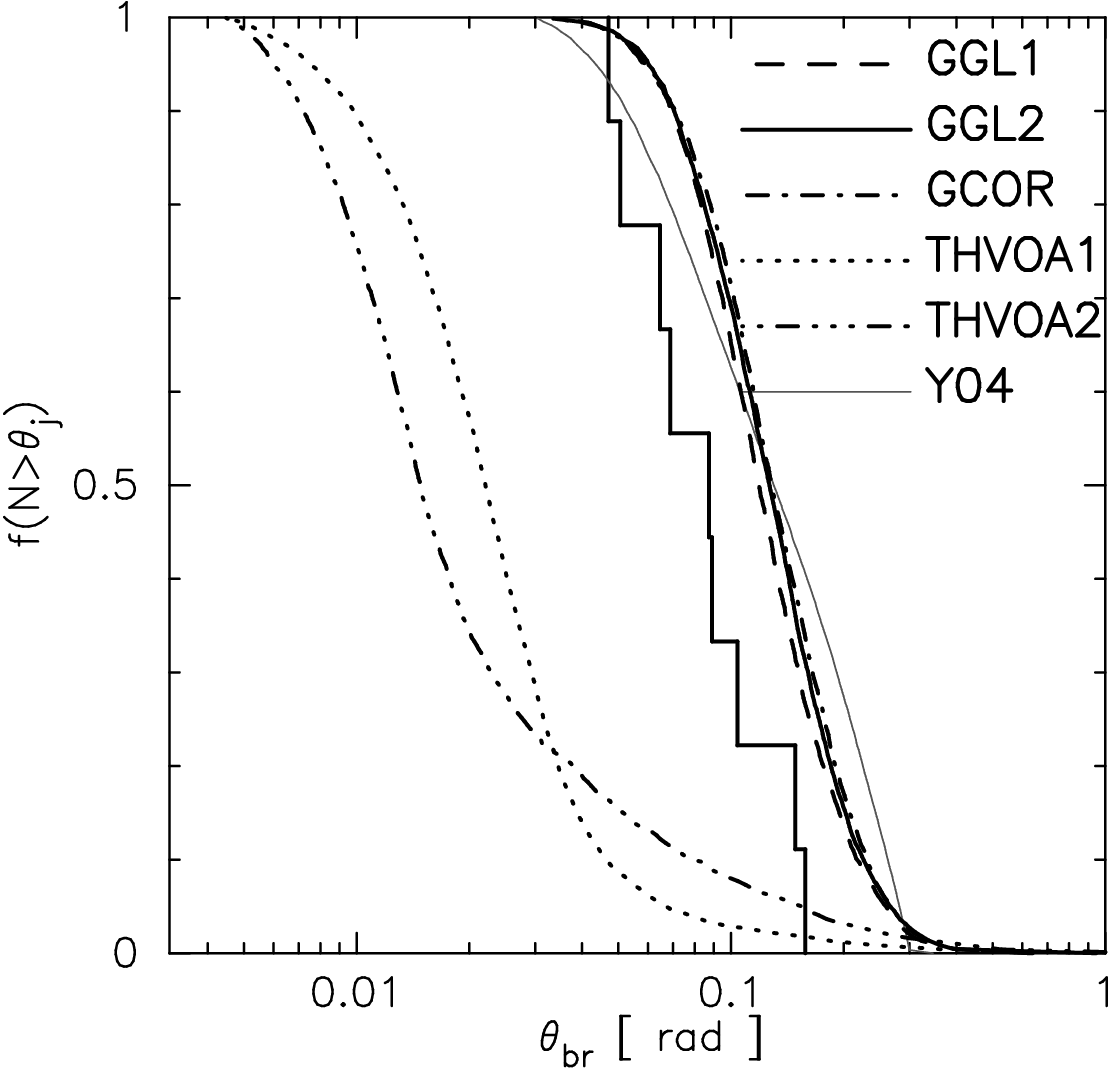}}}
\end{center}
\caption{The cumulative distribution in $\egi$ (left column) and
$\thetabr$ (right column) for detected bursts in the six models explored
in this paper.  The upper row compares models employing the GRBM
detector with data points observed by \bsaxnosp.  The lower row compares
models employing the WXM detector with data points observed by
\hetetwonosp.  The data used for the $\egi$ distribution are the same as
in Figure \ref{datasets}b, broken down by detector.  The data used in
the $\thetabr$ distribution is the dataset from \cite{ghirlanda2004}.}
\label{cuml_figs_EGI_theta}
\end{figure*}

\begin{figure*}[htb]
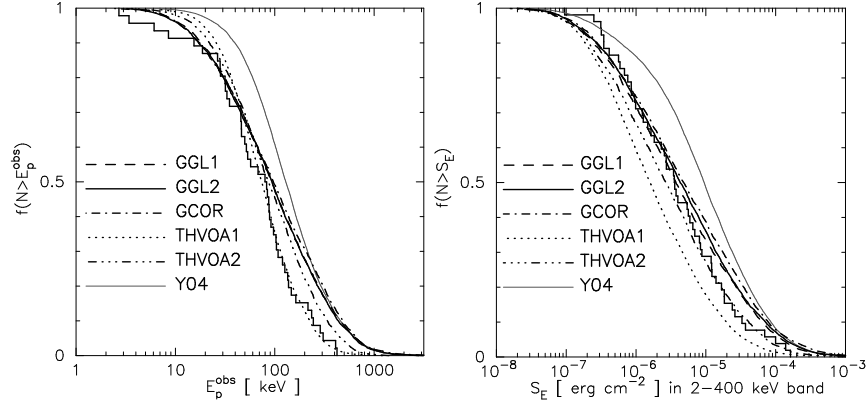

\begin{center}
\rotatebox{270}{\resizebox{5.3cm}{!}{\includegraphics{f13a_thin.eps}}}
\rotatebox{270}{\resizebox{5.3cm}{!}{\includegraphics{f13b_thin.eps}}}
\end{center}
\caption{The cumulative distribution in $\eop$ (left) and $S_{E}$(2-400
keV) (right) for detected bursts in the six models explored in this
paper.  These panels compare models employing the WXM detector with the
larger dataset of all bursts localized by \hetetwonosp. The data used for
these distributions are taken from \cite{sakamoto2005a} and
\cite{barraud2003}.}
\label{cuml_figs_hete}
\end{figure*}

Figures \ref{cuml_figs_Eiso_Ep}, \ref{cuml_figs_EGI_theta} and
\ref{cuml_figs_hete} summarize the six models by comparing the observed
data sets against the model cumulative distributions of $\eiso$, $\ep$,
$\egi$, $\thetabr$, $\eop$, and $S_{E}$(2-400 keV).  Although a great
deal of information is lost in projecting the 2-dimensional
distributions from the above figures onto each axis separately, some
useful information can still be obtained from these curves.  

Figure \ref{cuml_figs_Eiso_Ep} shows the cumulative distributions for
$\eiso$ and $\ep$, again separately comparing models using either the
WXM or GRBM instrumental threshold with data obtained from that
instrument.  Due to the small size of the current datasets, the THVOA,
GGL and GCOR models would not be judged as inconsistent with the data by
the KS test.  Model Y04 is here seen to over-produce brighter,
high-$\ep$ events, and under-produce low-$\ep$ XRFs.

Figure \ref{cuml_figs_EGI_theta} shows the cumulative distributions for
$\egi$ and $\thetabr$, again separately comparing models using either
the WXM or GRBM detector threshold with data obtained from that
detector.  This dataset is even sparser than that of Figure
\ref{cuml_figs_Eiso_Ep}, but the figure does highlight the effect of
rescaling the standard energy downward to $10^{49}$ ergs for the two
THVOA models.  The GGL and GCOR models avoid this problem by
incorporating the $\epeg$ relation, and thereby avoid the large
disparity with the $\egi$ and $\thetabr$ distributions seen for the
THVOA models.  

Figure \ref{cuml_figs_hete} shows the cumulative distributions for
$\eop$ and $S_{E}$(2-400 keV), comparing models using the WXM detector
threshold with the larger dataset of all bursts localized by
\hetetwonosp.  Again, the THVOA, GGL and GCOR models are all reasonably
consistent with the data.  There is some hint that the GGL and GCOR
models produce more bright, high-$\ep$ bursts than were seen by
\hetetwonosp, but this may be the result of trying to match both the
\bsax and \hetetwo populations with one model.

To summarize, models GGL2 and GCOR are the most successful at matching
the 2-dimensional distributions discussed above as well as the
individual cumulative distributions.  Model Y04, which seeks to explain
XRFs as off-jet GRBs, is unable to match the distributions of observed
XRFs.  The THVOA models produce large numbers of detectable bursts away
from the $\epei$ relation which are not seen in current datasets, and
require a re-scaled standard energy of $10^{49}$ ergs, which is
inconsistent with current afterglow theories \citep{panaitescu2001}.

\section{Conclusions}
\label{cha:conclude}

Bursts with known redshifts have been found to obey the $\epei$
relation, and a large population of bursts away from this relation is
not readily apparent in current datasets.  In particular, the limited
sample of XRFs with known redshift information is consistent with an
extension of the $\epei$ relation to over 5 orders of magnitude in
$\eiso$.  \cite{liang2004} found that the $\epei$ relation holds
internally within a large sample of bright BATSE bursts without known
redshift, perhaps indicating that the relation is a signature of the
physics of the emission mechanism.  However, recently some authors have
argued that the $\epei$ relation may be an artifact of some unknown
selection effect arising from the process of determining the burst
redshift, and that $25$\% \citep{nakar2004a} to $88$\% \citep{band2005}
of the BATSE bursts may be inconsistent with the $\epei$ relation. 
However, these results are controversial
\citep{ghirlanda2005a,bosnjak2005,lamb2005} and depend sensitively on
the quality of the spectral fit that generates the $\eop$ parameter.  I
therefore regard the question of the percentage of BATSE bursts that are
inconsistent with the $\epei$ relation to be an open one.

For low values of $\gamma$, top-hat variable opening-angle jet models
predict a sizable population of detectable, off-jet bursts that lie away
from the $\epei$ relation and that are not seen in current data sets. 
It may be that such bursts are in fact present in the BATSE catalog and
form a population of bursts that do not obey the $\epei$ relation.  On
the other hand, if such a population is found not to exist, it implies
that the bulk $\gamma$ of the jet is large ($\sim 300-1000$).  For
models that include the $\epeg$ relation, the off-jet burst population
lies closer to the $\epei$ relation than for other models, but a
similarly discrepant population of off-jet bursts is found to lie above
the $\epeg$ relation, leading to similar conclusions.

Regardless of the size of the population of off-jet bursts, it seems
unlikely that such a population can make up the bulk of the XRFs
observed by \hetetwonosp. The larger sample of \hetetwo bursts without
redshift information contains XRFs which lie toward smaller $\eop$
values than is predicted by the off-jet emission model and hence are not
easily explained as classical GRBs viewed off-jet.  Models in which XRFs
are the product of larger jet opening-angles seem to better match the
observed distributions of XRF properties.  This seems to match the
evidence arising from X-ray afterglows of XRFs.  \cite{granot2005}
calculated the afterglow light curves predicted by various models of
burst emission seen off-jet.  They find that a general feature of
off-jet afterglows is an initial rising light-curve that peaks at about
the jet break time and then declines rapidly, similar to an on-jet
event.  Afterglows with initially rising components have not been
observed.  In particular, XRFs with well-observed X-ray afterglows, for
example XRF 020427 \citep{amati2004} and XRF 050215b
\citep{sakamoto2005b}, have afterglow light curves that join smoothly
onto the end of the prompt emission and that show no evidence of a jet
break for many days after the burst, implying large jet opening-angles.

It is straightforward to arrange for top-hat variable opening-angle jet
models to match the empirical $\epeg$ relation, and such models also
provide a natural explanation for XRFs.  Figures \ref{fig_ggl1} and
\ref{fig_ggl2} illustrate the consequences of adopting the correlation
between $\ojet$ and $\egt$ that ensures that on-jet events obey the
$\epeg$ relation.  Most importantly, incorporating the $\epeg$ relation
in the THVOA model removes two of the main drawbacks of the THVOA model
presented in \cite{ldg2005}.  The requirement of very small ($\sim
2^{\circ}$) jet opening angles to explain the largest $\eiso$ values was
criticized \citep{stern2003} as being difficult to achieve in a
hydrodynamic jet.  The high end of the $\eiso$ distribution is here
explained by jets with moderate opening angles but larger $\egamma$
values.  The need to re-scale the central value of $\egt$ downward to
$\sim 10^{49}$ ergs to incorporate the XRFs in a unified model was
criticized as being difficult to reconcile with afterglow models.  The
$\epeg$ relation naturally produces a range of $\egamma$ values that
extends down into the XRF regime.

Matching the $\epeg$ relation also mitigates the problem of a large
population of bursts seen away from the $\epei$ relation in the low
$\gamma$ case.  In these models the off-jet events hew more closely to
the $\epei$ relation, but a similar problem arises in that these off-jet
events are seen away from the $\epeg$ relation instead.  A possible way
out of this dilemma might be to impose a relationship whereby narrower
jets have larger bulk $\gamma$ values and broader jets have smaller
$\gamma$ values.  Using a simple model where $\gamma \propto 1/\ojet$
results in a substantial reduction in the percentage of detected bursts
seen off-jet.

\cite{yamazaki2004b} have proposed a multiple subjet model for unifying
short and long GRBs, X-ray-rich GRBs and XRFs.  The model employs
emission from multiple subjets (seen off-subjet) to explain X-ray-rich
GRBs and XRFs.  The authors performed Monte Carlo simulations for a
universal multiple subjet model and find that the results are consistent
with the $\epei$ relation, albeit with considerable scatter (see Figure
4 in \cite{toma2005}).

There are two reasons why the multiple subjet model better satisfies the
$\epei$ relation.  First, the authors choose $\gamma = 300$, which
satisfies the constraint on $\gamma$ that I find above.  The behavior of
bursts viewed outside the envelope of subjets should approximate the
top-hat models considered in this work.  If the authors had adopted a
lower value of $\gamma$, the model would have produced a large number of
bursts that lie away from the $\epei$ relation; i.e., they would have
encountered the same problems as those of model THVOA1.  Second, since
$\gamma = 300$, the spectrum for each line of sight is dominated by that
of the closest subjet, and since there are many subjets, each line of
sight lies very close to at least one subjet, mitigating the effects of
relativistic kinematics produced by viewing a subjet well off the jet.

Finally, \cite{toma2005} find that for values of $\gamma$ lower than
$300$, the ratio between GRBs, X-ray-rich GRBs and XRFs becomes highly
skewed toward hard GRBs, in contradiction with the HETE-2 results. 
Thus, all of the results in \cite{toma2005} support the requirement I
find in this paper that large gamma values are needed in order to match
the observed data for XRFs, X-ray-rich GRBs, and GRBs.

Off-jet relativistic kinematics will be important in non-uniform jets as
well as in top-hat jets.  Models employing Gaussian \citep{zhang2004} or
Fisher-shaped \citep{dlg2005a} jets rely on the exponential fall off of
the emissivity with viewing angle to match the wide spread of observed
burst quantities.  By including off-jet emission in these models, the
exponential fall off will be dominated at some angle by the power-law
fall off due to relativistic kinematics, thereby broadening the
emissivity distribution and reducing the range of generated $\eiso$
values.  \cite{graziani2005} showed that different underlying burst
profiles may have radically different observational distributions.  We
hope to use population synthesis Monte Carlo simulations to further
explore these models in future work.

In conclusion, off-jet emission from collimated GRB outflows should
exist simply as a consequence of relativistic kinematics.  Monte Carlo
population synthesis simulations of top-hat shaped variable
opening-angle jet models predict a large population of off-jet bursts
that are observable and that lie away from the $\epei$ and $\epeg$
relations. Such off-jet events are not apparent in current datasets. 
These discrepancies can be removed if $\gamma > 300$ for all bursts or
if there is a strong inverse correlation between $\gamma$ and $\ojet$. 
The simulations show that XRFs seen by \hetetwo and \bsax cannot be
easily explained as classical GRBs viewed off-jet, and are more
naturally explained as jets with large opening-angles.

\acknowledgments

Many thanks are due to Don Lamb and Carlo Graziani for their invaluable
advice and assistance in completing this project.  I would also like to
thank Nat Butler and Takanori Sakamoto, for stimulating and helpful
conversations about spectral analysis and the nature of off-jet
emission.


\begin{thebibliography}{999}
 
\bibitem[Amati et al.(2002)]{amati2002}
	Amati, L., et al. 2002, \aap, 390, 81
\bibitem[Amati et al.(2004)]{amati2004}
	Amati, L., et al. 2004, \aap, 426, 415
\bibitem[Band et al.(1993)]{band1993}
	Band, D. L. et al. 1993, \apj, 413, 281
\bibitem[Band(2003)]{band2003}
	Band, D. L. 2003, \apj, 588, 945
\bibitem[Band \& Preece(2005)]{band2005}
	Band, D. L. \& Preece, R. D. 2005, \apj, 627, 319 
\bibitem[Barraud et al.(2003)]{barraud2003}
	Barraud, C., et al. 2003, \aap, 400, 1021
\bibitem[Bloom et al.(2003)]{bloom2003}
	Bloom, J., Frail, D. A., \& Kulkarni, S. R. 2003, \apj, 588, 945
\bibitem[Bosnjak et al.(2005)]{bosnjak2005}
	Bosnjak, Z., et al. 2005, MNRAS, submitted (astro-ph/0502185)
\bibitem[Donaghy(2005a)]{donaghy2005a}  
    Donaghy, T. Q. 2005a, Il Nuovo Cimento, 28C, 407
\bibitem[Donaghy, Lamb, \& Graziani(2005a)]{dlg2005a}  
    Donaghy, T. Q., Lamb, D. Q., \& Graziani, C. 2005a, Il Nuovo
	Cimento, 28C, 403	
\bibitem[Donaghy, Lamb, \& Graziani(2005b)]{dlg2005b}  
	Donaghy, T. Q., Lamb, D. Q., \& Graziani, C. 2005b, \apj, submitted 
\bibitem[Donaghy, Lamb, \& Graziani(2005c)]{dlg2005c}  
	Donaghy, T. Q., Lamb, D. Q., \& Graziani, C. 2005c, \apj, submitted 
\bibitem[Frail et al.(2001)]{frail2001}
	Frail, D., et al. 2001, \apj, 562, L55
\bibitem[Friedman \& Bloom(2004)]{friedman2004}
	Friedman, A. S. \& Bloom, J. S. 2004, \apj, 627, 1 
\bibitem[Ghirlanda et al.(2004)]{ghirlanda2004}
	Ghirlanda, G., Ghisellini, G. \& Lazzati, D. 2004, \apj, 616, 331
\bibitem[Ghirlanda et al.(2005a)]{ghirlanda2005a}
	Ghirlanda, G., Ghisellini, G., \& Firmani, C. 2005a, 
	MNRAS, 361, L10 
\bibitem[Ghirlanda et al.(2005b)]{ghirlanda2005b}
	Ghirlanda, G., Ghisellini, G., Lazzati, D. \& Firmani, C. 2005b, 
	Il Nuovo Cimento, 28C, 303	
\bibitem[Golenetskii et al.(2005a)]{golenetskii2005a}
	Golenetskii, S., et al. 2005a, GCN Circ 3179, 
	http://gcn.gsfc.nasa.gov/gcn/gcn3/3179.gcn3
\bibitem[Golenetskii et al.(2005b)]{golenetskii2005b}
	Golenetskii, S., et al. 2005b, GCN Circ 3473,
	http://gcn.gsfc.nasa.gov/gcn/gcn3/3473.gcn3
\bibitem[Golenetskii et al.(2005c)]{golenetskii2005c}
	Golenetskii, S., et al. 2005c, GCN Circ 3518,
	http://gcn.gsfc.nasa.gov/gcn/gcn3/3518.gcn3
\bibitem[Granot et al.(2005)]{granot2005}
	Granot, J., Ramirez-Ruiz, E. \& Perna, R. 2005, \apj, 630, 1003 
\bibitem[Graziani et al.(2005)]{graziani2005}
	Graziani, C., Donaghy, T. Q., \& Lamb, D. Q. 2005, \apj, in press
	(astro-ph/0505623)
\bibitem[Harrison et al.(1999)]{harrison1999}
	Harrison, F. A., et al. 1999, \apj, 559, 123
\bibitem[Heise et al.(2001)]{heise2000}
	Heise, J., in't Zand, J., Kippen, R. M., \& Woods, P. M., in
	Proc. 2nd Rome Workshop:  Gamma-Ray Bursts in the Afterglow
	Era, eds. E. Costa, F. Frontera, J. Hjorth (Berlin:
	Springer-Verlag), 16
\bibitem[Ioka \& Nakamura(2001)]{ioka2001}
	Ioka K. \& Nakamura T. 2001, \apj, 554, L163
\bibitem[Kippen et al.(2001)]{kippen2002}
	Kippen, R. M., Woods, P. M., Heise, J., in't Zand, J., Briggs,
	M.S., \& Preece, R. D. 2003, in AIP Conf. Proc. 662, Gamma-Ray Burst
	and Afterglow Astronomy, ed. G. R. Ricker \& R. K.
	Vanderspek (New York: AIP), 244
\bibitem[Kulkarni et al.(1998)]{kulkarni1998}
	Kulkarni, S. R., et al. 1998, Nature, 393, 35
\bibitem[Kulkarni et al.(1999)]{kulkarni1999}
	Kulkarni, S. R., et al. 1999, Nature, 398, 389
\bibitem[Lamb(1999)]{lamb1999}
	Lamb, D. Q. 1999, A\&AS, 138, 607
\bibitem[Lamb(2000)]{lamb2000}
	Lamb, D. Q. 2000, Phys. Rep., 333, 505
\bibitem[Lamb, Donaghy, \& Graziani(2005)]{ldg2005}  
	Lamb, D. Q., Donaghy, T. Q., \& Graziani, C. 2004, \apj, 620, 355
\bibitem[Lamb et al.(2005)]{lamb2005}
	Lamb, D. Q., et al. 2005, \apj, submitted  
\bibitem[Liang, Dai \& Wu(2004)]{liang2004}
	Liang, E. W., Dai, Z. G., \& Wu, X. F. 2004, \apj, 606, L29
\bibitem[Lloyd-Ronning et al.(2000)]{lloyd-ronning2000}
	Lloyd-Ronning, N., Petrosian, V., \& Mallozzi, R. S. 2000, \apj, 534, 227
\bibitem[Nakar \& Piran(2004)]{nakar2004a}
	Nakar, E. \& Piran, T. 2004, \apj, submitted (astro-ph/0412232)
\bibitem[Panaitescu \& Kumar(2001)]{panaitescu2001}
	Panaitescu, A. \& Kumar, P. 2001, \apj, 560, L49
\bibitem[Rhoads(1997)]{rhoads1997}
	Rhoads, J. E. 1997, \apj, 478, L1
\bibitem[Ricker et al.(2001)]{ricker2003}
	Ricker, G. R., et al. 2003, in AIP Conf. Proc. 662, Gamma-Ray Burst
	and Afterglow Astronomy, ed. G. R. Ricker \& R. K.
	Vanderspek (New York: AIP), 3
\bibitem[Rowan-Robinson(2001)]{rr2001}
	Rowan-Robinson M. 2001, \apj, 549, 745
\bibitem[Sakamoto et al.(2004)]{sakamoto2004}
	Sakamoto, T., et al. 2004, \apj, 602, 875
\bibitem[Sakamoto et al.(2005a)]{sakamoto2005a}
	Sakamoto, T., et al. 2005a, \apj, 629, 311
\bibitem[Sakamoto et al.(2005b)]{sakamoto2005b}
	Sakamoto, T., et al. 2005b, \apj, submitted  
\bibitem[Sari, Piran \& Halpern(1999)]{sph1999}
	Sari, R., Piran, T., Halpern, J. P. 1999, \apj, 519, L17
\bibitem[Stern(2003)]{stern2003}
	Stern, B. E. 2003, MNRAS, 345, 590
\bibitem[Toma et al.(2005)]{toma2005}
	Toma, K., Yamazaki, R., \& Nakamura T. 2005, \apj, 635, 481
\bibitem[Yamazaki, Ioka \& Nakamura(2002)]{yamazaki2002}
	Yamazaki, R., Ioka K., \& Nakamura T. 2002, \apj, 571, L31
\bibitem[Yamazaki, Ioka \& Nakamura(2003)]{yamazaki2003}
	Yamazaki, R., Ioka K., \& Nakamura T. 2003, \apj, 593, 941
\bibitem[Yamazaki, Ioka \& Nakamura(2004a)]{yamazaki2004a}
	Yamazaki, R., Ioka K., \& Nakamura T. 2004a, \apj, 606, L33
\bibitem[Yamazaki, Ioka \& Nakamura(2004b)]{yamazaki2004b}
	Yamazaki, R., Ioka K., \& Nakamura T. 2004b, \apj, 607, L103
\bibitem[Zhang et al.(2004)]{zhang2004}
	Zhang, B., Dai, X., Lloyd-Ronning, N. M., \& M\'esz\'aros, P.
	2004, 601, L119

\end{thebibliography}
\end{document}